\documentclass[twocolumn, twocolappendix]{aastex631}

\usepackage{booktabs}
\usepackage{array}
\usepackage{multirow}
\usepackage{float}
\usepackage{lipsum}
\usepackage{enumitem}

\newcolumntype{x}[1]{>{\centering\arraybackslash\hspace{0pt}}p{#1}}

\makeatletter
\let\frontmatter@title@above=\relax
\makeatother

\newif\ifedits
\editsfalse
\ifedits
    \renewcommand{\edit}[1]{{\color{red} #1}}
    
\else
    \renewcommand{\edit}[1]{#1}
    
\fi

\begin{document}

\title{Metallicity Gradients in Modern Cosmological Simulations II: The Role of Bursty Versus Smooth Feedback at High-Redshift}

\shorttitle{The Role of Bursty Versus Smooth Feedback on Metallicity Gradients}
\shortauthors{Garcia et al.}

\correspondingauthor{Alex M. Garcia}
\email{alexgarcia@virginia.edu}

\author[0000-0002-8111-9884]{Alex M. Garcia}
\affiliation{Department of Astronomy, University of Virginia,
530 McCormick Road, Charlottesville, VA 22904}
\affiliation{Virginia Institute for Theoretical Astronomy, University of Virginia, Charlottesville, VA 22904, USA}
\affiliation{The NSF-Simons AI Institute for Cosmic Origins, USA}

\author[0000-0002-5653-0786]{Paul Torrey}
\affiliation{Department of Astronomy, University of Virginia, 
530 McCormick Road, 
Charlottesville, VA 22904}
\affiliation{Virginia Institute for Theoretical Astronomy, University of Virginia, Charlottesville, VA 22904, USA}
\affiliation{The NSF-Simons AI Institute for Cosmic Origins, USA}

\author[0000-0003-0275-5506]{Aniket Bhagwat}
\affiliation{Max Planck Institut f\"ur Astrophysik, Karl Schwarzschild Stra{\ss}e 1, D-85741 Garching, Germany}

\author[0000-0002-6196-823X]{Xuejian Shen}
\affiliation{Department of Physics and Kavli Institute for Astrophysics and Space Research, Massachusetts Institute of Technology, Cambridge, MA 02139, USA}

\author[0000-0001-8593-7692]{Mark Vogelsberger}
\affil{Department of Physics and Kavli Institute for Astrophysics and Space Research, Massachusetts Institute of Technology, Cambridge, MA 02139, USA}
\affil{Fachbereich Physik, Philipps Universit\"at Marburg, D-35032 Marburg, Germany}


\author[0009-0009-5565-3790]{William McClymont}
\affiliation{Kavli Institute for Cosmology, University of Cambridge, Madingley Road, Cambridge CB3 0HA, UK}
\affiliation{Cavendish Laboratory, University of Cambridge, 19 JJ Thomson Avenue, Cambridge CB3 0HE, UK}

\author[0009-0002-6248-3688]{Jaya Nagarajan-Swenson}
\affiliation{Department of Astronomy, University of Virginia,
530 McCormick Road, Charlottesville, VA 22904}

\author[0000-0003-3569-4092]{Sophia G. Ridolfo}
\affiliation{Harvard-Smithsonian Center for Astrophysics, 60 Garden Street, Cambridge, MA 02138, USA}

\author[0000-0002-1333-147X]{Peixin Zhu}
\affiliation{Harvard-Smithsonian Center for Astrophysics, 60 Garden Street, Cambridge, MA 02138, USA}

\author[0009-0008-7017-5742]{Dhruv T. Zimmerman}
\affiliation{Department of Astronomy, University of Florida, 211 Bryant Space Sciences Center, Gainesville, FL 32611 USA}

\author[0000-0003-1811-8915]{Oliver Zier}
\affiliation{Harvard-Smithsonian Center for Astrophysics, 60 Garden Street, Cambridge, MA 02138, USA}

\author[0000-0002-4826-8642]{Sarah Biddle}
\affiliation{Harvard-Smithsonian Center for Astrophysics, 60 Garden Street, Cambridge, MA 02138, USA}

\author[0000-0002-5222-1337]{Arnab Sarkar}
\affiliation{Department of Physics, University of Arkansas, 825 W Dickson Street, Fayetteville, AR, 72701}
\affiliation{Kavli Institute for Astrophysics and Space Research, Massachusetts Institute of Technology, 70 Vassar Street, Cambridge, MA 02139, USA}

\author[0000-0002-4469-2518]{Priyanka Chakraborty}
\affiliation{Department of Physics, University of Arkansas, 825 W Dickson Street, Fayetteville, AR, 72701}
\affiliation{Harvard-Smithsonian Center for Astrophysics, 60 Garden Street, Cambridge, MA 02138, USA}

\author[0000-0002-1367-0949]{Ruby J. Wright}
\affiliation{International Centre for Radio Astronomy Research (ICRAR), M468, University of Western Australia, 35 Stirling Hwy, Crawley, WA 6009, Australia}

\author[0000-0002-3247-5321]{Kathryn~Grasha}
\altaffiliation{ARC DECRA Fellow}
\affiliation{Research School of Astronomy and Astrophysics, Australian National University, Canberra, ACT 2611, Australia}

\author[0000-0002-6748-2900]{Tiago Costa}
\affiliation{School of Mathematics, Statistics and Physics, Newcastle University, Newcastle upon Tyne, NE1 7RU, UK}

\author[0000-0001-5211-1958]{Laura Keating}
\affiliation{Institute for Astronomy, University of Edinburgh, Blackford Hill, Edinburgh, EH9 3HJ, UK}

\author[0000-0001-6092-2187]{Rahul Kannan}
\affiliation{Department of Physics and Astronomy, York University, 4700 Keele Street, Toronto, ON M3J 1P3, Canada}

\author[0000-0002-2838-9033]{Aaron Smith}
\affiliation{Department of Physics, The University of Texas at Dallas, Richardson, TX 75080, USA}

\author[0000-0002-6021-7020]{Enrico Garaldi}
\affiliation{Kavli IPMU (WPI), UTIAS, The University of Tokyo, Kashiwa, Chiba 277-8583, Japan}

\author[0000-0001-8778-7587]{Ewald Puchwein}
\affiliation{Leibniz-Institut f\"ur Astrophysik Potsdam, An der Sternwarte 16, 14482 Potsdam, Germany}

\author{Benedetta Ciardi}
\affiliation{Max Planck Institut f\"ur Astrophysik, Karl Schwarzschild Stra{\ss}e 1, D-85741 Garching, Germany}

\author[0000-0001-6950-1629]{Lars Hernquist}
\affiliation{Harvard-Smithsonian Center for Astrophysics, 60 Garden Street, Cambridge, MA 02138, USA}

\author[0000-0001-8152-3943]{Lisa J. Kewley}
\affiliation{Harvard-Smithsonian Center for Astrophysics, 60 Garden Street, Cambridge, MA 02138, USA}

\begin{abstract}
The distribution of gas-phase metals within galaxies encodes the impact of stellar feedback on galactic evolution.
At high-redshift, when galaxies are rapidly assembling, feedback-driven outflows and turbulence can strongly reshape radial metallicity gradients.
In this work, we use the FIRE-2, SPICE, Thesan and Thesan Zoom cosmological simulations -- spanning a range of stellar feedback from bursty (time-variable) to smooth (steady) -- to investigate how these feedback modes shape gas-phase metallicity gradients at $3<z\lesssim11$.
Across all models, we find that galaxies with bursty feedback (FIRE-2, SPICE Bursty, and Thesan Zoom) develop systematically flatter (factors of $\sim2-10$) metallicity gradients than those with smooth feedback (SPICE Smooth and Thesan Box), particularly at stellar masses $M_\star > 10^{9}~{\rm M_\odot}$.
These results demonstrate that bursty stellar feedback provides sufficient turbulence to prevent strong negative gradients from forming, while smooth stellar feedback does not generically allow for efficient radial redistribution of metals thereby keeping gradients steep.
Finally, we compare with recent observations, finding that the majority -- but, notably, not all -- of the observed gradients may favor a bursty stellar feedback scenario.
In all, these results highlight the utility of high-resolution observations of gas-phase metallicity at high-redshift as a key discriminator of these qualitatively different feedback types.
\end{abstract}


\section{Introduction}

Feedback is central to galaxy evolution, but capturing its impact on galaxies remains a major challenge for simulations.
Stars are among the most important sources of feedback, driving it through multiple channels including supernovae, photoionization and radiation pressure, protostellar jets, and stellar winds \citep[e.g.,][]{Jijina_1996,Yorke_1989,Dale_2005,Quillen_2005,Evans_2009,Lopez_2011}.
These processes can influence properties on the galactic scales, yet modern galaxy evolution simulations lack the resolution to directly model the micro-physics of stellar feedback and turbulence within the \edit{interstellar medium (ISM;} \citeauthor{Crain_2023}~\citeyear{Crain_2023}).
Nonetheless, accurately modeling feedback on these unresolved scales is imperative to producing realistic galaxy populations.

At present, galaxy evolutionary simulations that model baryonic feedback can be broadly categorized as either: 
{\it (i)} high-resolution zoom-in simulations of individual galaxies and their local environment ($\sim$a few Mpc scales) and
{\it (ii)} lower-resolution large ($\gtrsim35-300$ Mpc scales) cosmological box simulations \citep[see][for a review]{Vogelsberger_2020}.

High-resolution simulations (e.g. FIRE; \citeauthor{Hopkins_2014} \citeyear{Hopkins_2014}, or SMUGGLE; \citeauthor{Marinacci_2019} \citeyear{Marinacci_2019}) can begin to resolve the multiphase ISM down to approximately giant molecular cloud ($10^4~{\rm M}_\odot$) scales, allowing for explicit modeling of star formation, feedback, and cooling.
Such simulations have been successful at reproducing a number of galactic-scale properties \citep[][etc.]{El_Badry_2016,Sparre_2017,Ma_2017,Torrey_2017}.
However, owing to the high computational cost, it is difficult to produce large samples of galaxies across a variety of cosmological environments (although, see recent efforts by FIREbox, \citeauthor{Feldmann_2023} \citeyear{Feldmann_2023}, and COLIBRE, \citeauthor{Schaye_2025} \citeyear{Schaye_2025}).
Lower-resolution large-box simulations, on the other hand, provide a large population of galaxies in a wide diversity of environments \citep[][etc.]{Vogelsberger_2014a, Schaye_2015, Dave_2019, Pillepich_2018a}.
The cosmological box simulations have had their share of success reproducing population-level scaling relationships \citep[e.g.,][]{Agertz_2011,Genel_2014,Sparre_2015,Ferrero_2017,Dave_2019,Torrey_2019}, but struggle at high-redshift and small scales \citep{Sun_2023,Shen_2024b,Qi_2025}.
The fundamental limitation of these models is that the scales of star formation and feedback are entirely unresolved.
These processes are therefore usually treated entirely as phenomenological prescriptions.
Typical treatments rely on simplifying assumptions, such as describing the behavior of the dense, star-forming ISM with an effective equation of state \citep[e.g.,][]{Springel_Hernquist_2003,Schaye_DallaVechhia_2008} or temporarily decoupling feedback-driven winds from the hydrodynamics to help facilitate the escape of gas \citep{Vogelsberger_2013,Dave_2019}.

\begin{figure}
    \centering
    \includegraphics[width=\linewidth]{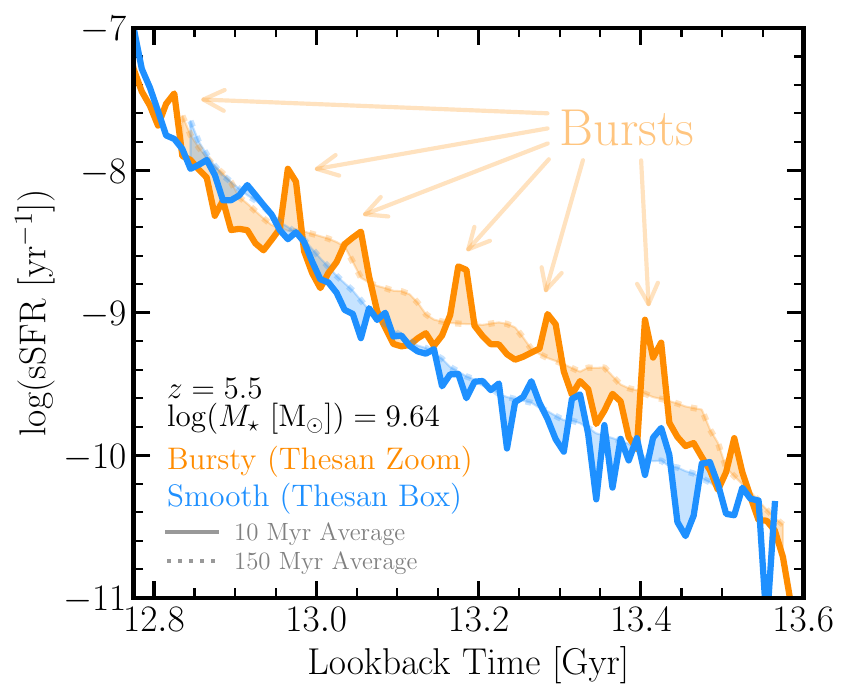}
    \caption{{\bf Example (Specific) Star Formation Histories in Bursty and Smooth Feedback Models.}
    The mass assembly of two galaxies with stellar mass $\log(M_\star~[{\rm M}_\odot])=9.64$ at $z\sim5.5$ from Thesan Zoom (bursty; orange) and Thesan Box (smooth; blue).
    The solid lines are the specific star formation rate (sSFR) of the galaxy averaged over $10$ Myr, the \edit{dotted} lines represent the sSFR averaged over $150$ Myr, and the shaded region represents the difference between the two.
    The bursty model has episodic factors of a few increases in star formation rates which lead to bursts of feedback.
    While the ``smooth'' model galaxy does still have time variations of their sSFR, particularly at very early times ($\gtrsim13.3$ Gyr lookback time), it is generally much less.
    Finally, we note that these two galaxies are not the same halo run with different models and were instead selected to have the same stellar mass at $z=5.5$.
    }
    \label{fig:SFH_plot}
\end{figure}

A key consequence of different treatments of star formation and feedback is that the star formation histories of galaxies in explicit models are much more bursty (strongly time variable) than those of equation of state ISM models.
The bursty nature of the feedback from explicit models comes from their self-regulation of the star-forming ISM.
While an equation of state provides additional pressure support that prevents the collapse of all gas into stars \citep{Springel_Hernquist_2003,Schaye_DallaVechhia_2008,Ploeckinger_2024,Ploeckinger_2025,Burger_2025}, such pressure support is not included out-of-the-box in explicit ISM models.
Therefore, the treatment of feedback needs to prevent all of the local gas from converting into stars.
This explicit self-regulation by feedback from stars leads to an episodic nature of star formation followed by rapid quenching.

To put this more concretely, we show two example mass assembly histories in Figure~\ref{fig:SFH_plot} for two galaxies of the same stellar mass at $z\sim5.5$ in Thesan Box (Smooth) and Thesan Zoom project (Bursty; simulations described in more detail in Section~\ref{subsubsec:thesan}).
The explicit ISM model (labeled bursty) has several bursts of star formation (rapid enhancements and decreases of factors of a few; see also \citeauthor{McClymont_2025c} \citeyear{McClymont_2025c}), while the equation of state ISM (labeled smooth) model has a much smoother star formation history.\ignorespaces
\footnote{\ignorespaces
It should be noted that the ``smooth'' star formation does still have time variation; however, it is on the tens of percent level for this system, unlike the factors of several in its bursty counterpart.
}
Bursts of star formation generally lead to episodic blow-outs of gas as large populations of massive stars formed during the burst die concurrently, thereby preventing further local star formation
\citep{Muratov_2015,Muratov_2017,Angles_Alcazar_2017b,Pandya_2021}.

One advantage of bursty feedback is that it can naturally be invoked to alleviate small-scale tensions with $\Lambda$CDM, such as the core-cusp problem \cite[][]{Read_2005,Pontzen_2012,Lazar_2020,Mostow_2024} by driving rapid fluctuations of the gravitational potential, and the overabundance of UV bright galaxies at high-redshift \citep[e.g.,][]{Mason_2023,Shen_2023,Shen_2024,Gelli_2024,Sun_2023,Narayanan_2024} with bursts contributing to a high fraction of the UV luminosity.
Bursts are therefore not just a unique modeling quirk of high-resolution simulations, but a potentially necessary ingredient in galaxy evolution \citep[see also][etc]{Looser_2024,Danhaive_2025,McClymont_2025b,McClymont_2025,Witten_2025}.
It is not fully clear at present, however, the impacts of these systematic and episodic blowouts are on the rest of the galaxy.

A method by which the effect of rapid, large outflows triggered by star formation can be quantified is by looking at the overall gas-phase metal content, or metallicity, of the galaxy \citep[see, e.g., work by][]{Garcia_2023,Garcia_2024a,Garcia_2024b,Garcia_2025a,Garcia_2025b,Bassini_2024,Marszewski_2025,McClymont_2025}.
Metals are ideal as they produce bright emission lines that can be easily observed in the (rest) optical \citep{Kewley_2019,Maiolino_Mannucci_2019} while tracing the flows and chemical enrichment of the system \citep{Tumlinson_2017,Muratov_2017,Peroux_2020}.
Metallicity thus offers an observable, {\it direct} probe of how feedback regulates galaxies' growth and evolution.
The spatial distribution of metals within galaxies traces the system's assembly and evolution, making it a particularly useful tool.

Local galaxies generally exhibit higher metallicities in their centers than their outskirts, a pattern usually characterized by a negative (radially decreasing) metallicity gradient \citep{Searle_1971,Sanchez_2013,Stanghellini_2015,Franchetto_2021,Nelson_2021}.
The existence of negative gradients is usually explained with inside-out galaxy growth, wherein the stellar populations in the center of the galaxy form and evolve earlier than on the outskirts, chemically enriching the inner regions earlier \citep[e.g.,][]{Prantzos_2000,Pilkington_2012,Perez_2013,Tissera_2019}.
The characteristic negative gradients can be flattened (perhaps even inverted) by radial mixing within the disk via winds \citep{Gibson_2013,Ma_2017,Garcia_2023}, gas flows along spiral arms \citep{Friedli_1994,Friedli_1995,Orr_2023}, mergers \citep{Rupke_2010a,Rupke_2010b,Torrey_2012}, or through pristine gas inflows into the central regions of galaxies \citep{Ceverino_2016,Tissera_2022,Rodriguez_Del_Pino_2024,Tapia_2025}.
Moreover, the host mass of a galaxy has been shown to be important for metallicity gradients, both in simulations and observations \citep{Belfiore_2017,Garcia_2025b}.
As such, a wide variety of negative, flat, and positive gradients are observed across time \citep[e.g.,][]{Wuyts_2016,Wang_2017,Wang_2019,Wang_2020,Curti_2020,Simons_2021,Grasha_2022,Vale_2025}, although observational systematics (e.g., limited angular resolution, contributions from diffuse ionized gas) can also have systematic impacts on measured gradients \citep[see, e.g.,][]{Yuan_2013,Poetrodjojo_2019,Acharyya_2020}.

In \cite{Garcia_2025b}, we show that modern large volume cosmological simulations are in tension with high-redshift ($z>3$) observations of metallicity gradients \citep{Troncoso_2014,Wang_2022,Arribas_2024,Tripodi_2024,Vallini_2024,Venturi_2024}.
We find that EAGLE \citep{Schaye_2015}, Illustris \citep{Vogelsberger_2014a}, IllustrisTNG \citep{Pillepich_2018a}, and SIMBA \citep{Dave_2019}---notably all effective equation of state ISM models---produce systematically stronger gradients (i.e., more negative) than are observed at $z>3$, especially in intermediate-to-high mass galaxies ($10^9~{\rm M}_\odot < M_\star<10^{10}~{\rm M}_\odot$), suggesting that these models do not mix their gas content sufficiently.
Since that work, there have been a few more gradients measured at high-redshift with {\it JWST} at $z\gtrsim3$ \citep{Acharyya_2025,Ivey_2025,Li_2025}.
While \cite{Li_2025} presents some evidence for strong negative gradients at high-redshift, the preference for flat gradients at high-redshift persists.
If galaxies truly do have systematically flat gradients at high-redshift, then many modern cosmological simulations do not accurately model galaxy formation in its earliest stages.

Stellar feedback plays a central role in the evolution of metallicity gradients.
\cite{Gibson_2013} shows, using the MUGS \citep{Stinson_2010} and MaGICC \citep{Brook_2012} models, that feedback variations can lead to qualitatively different behavior of metallicity gradients.
They find that the enhanced feedback models of MaGICC lead to systematically flatter gradients than the more conservative MUGS model.
Inspired by the results of \cite{Gibson_2013}---and continuing the effort started in \cite{Garcia_2025b}---we aim here to quantify the role of {\it bursty} feedback in resolving the metallicity gradient tension at high-redshift.

The rest of this paper is arranged as follows.
In Section~\ref{sec:methods}, we introduce the simulation models we analyze in this work, define our galaxy selection criteria, and then describe our methodology for calculating metallicity gradients.
In Section~\ref{sec:results}, we show the full distribution of high-redshift ($3<z\lesssim11$) gradients in each simulation and then break our samples into both stellar mass and redshift bins.
In Section~\ref{sec:discussion}, we contextualize the results of this work and \cite{Garcia_2025b} to present a complete picture of the gas-phase metallicity gradient evolution over 13 billion years in cosmological simulations and compare to recent observational results at high-redshift ($z>3$).
Finally, in Section~\ref{sec:conclusions}, we state our conclusions.

\section{Methods}
\label{sec:methods}

\defcitealias{Springel_Hernquist_2003}{SH03}
\defcitealias{Hopkins_2018}{H18}
\defcitealias{Marinacci_2019}{M19}
\defcitealias{Kimm_Cen_2014}{KC14}
\begin{table*}
    \centering
    \renewcommand{\arraystretch}{1.25}
    \begin{tabular}{m{0.2\linewidth}x{0.17\linewidth}x{0.17\linewidth}x{0.17\linewidth}x{0.17\linewidth}}
        \toprule
        & {FIRE-2}  & {SPICE} & {Thesan Box} & {Thesan Zoom}
        \\\midrule\midrule 
        Classification & Bursty & Has Both & Smooth & Bursty \\
        Box Length~[${\rm cMpc}\,h^{-1}$] & --- & $10$ & $64.7$ & --- \\
        $m_{\rm baryon}$~[${\rm M}_\odot$] & $119-7100$ & $975$ & $3.12\times10^6$ & $142-9090$\\
        Feedback Mechanisms & Stars & Stars & Stars and AGN & Stars \\
        ISM Model & FIRE-2 (\protect\citetalias{Hopkins_2018}) & Turbulence regulated & eEOS (\citetalias{Springel_Hernquist_2003}) & SMUGGLE (\protect\citetalias{Marinacci_2019}) \\
        Stellar Feedback & Radiative Feedback + SNe + winds & Mechanical SNe & Thermal SNe + Kinectic winds & Radiation Pressure + SNe + winds \\
        AGN Feedback & --- & --- & Thermal + Kinetic & --- \\
        SF Criterion & Self-grav. collapse & Turbulence regulated & Density Threshold & Self-grav. collapse \\
        SF Density~[${\rm cm}^{-3}$] & $1000$ & $10$ & $0.1$ & $10$\\
        SF Efficiency & Feedback regulated & Variable $\epsilon_{\rm ff}$ & Constant $\epsilon_{\rm ff}$ & Feedback regulated \\
        Code & {\sc gizmo} & {\sc ramses} & {\sc arepo} & {\sc arepo}\\
        Model Reference & \protect\cite{Hopkins_2018} & {\protect\cite{Bhagwat_2024}} & \protect\cite{Kannan_2022} & \protect\cite{Kannan_2025}\\
        \bottomrule
    \end{tabular}
    \caption{\edit{{\bf Summary of Simulations.}
    A brief summary of the simulation models used in this work. 
    We list the general stellar feedback type (smooth or bursty), size of the volume (when applicable), initial baryon mass resolution ($m_{\rm baryon}$), list the relevant feedback mechanisms of the models (and details appertaining), and provide references for the models~(noting that several of the models have multiple references, see Section~\ref{subsec:simulations} for a more complete accounting).
    We note that both the SPICE Smooth and Bursty models have the same parameters for all table entries, save for their feedback type.
    }
    }
    \label{tab:simulations}
\end{table*}

We employ data products from the FIRE, SPICE, and Thesan cosmological simulations.
The advantage of these models for this work is that they offer an assortment of physical implementations, both bursty and smooth stellar feedback. 
Moreover, each of these models includes simulations of massive galaxies ($M_\star \gtrsim 10^8~{\rm M}_\odot$) in the early universe ($z \gtrsim 5$), allowing us to probe the earliest onset of galactic evolution.
\edit{We summarize the main features of each model in Table~\ref{tab:simulations}.}

Critically, the models of this work can be broadly categorized as either having bursty or smooth stellar feedback.
FIRE, SPICE Bursty, and Thesan Zoom have bursty stellar feedback, while SPICE Smooth and Thesan Box have smooth stellar feedback.
In this Section, we introduce the simulation models (\S\ref{subsec:simulations}), detail our selection criteria (\S\ref{subsec:selection}), and then define our method for obtaining the gas-phase metallicity gradients from each galaxy (\S\ref{subsec:gradient_definitions}).

\begin{figure*}
    \centering
    \includegraphics[width=0.95\linewidth]{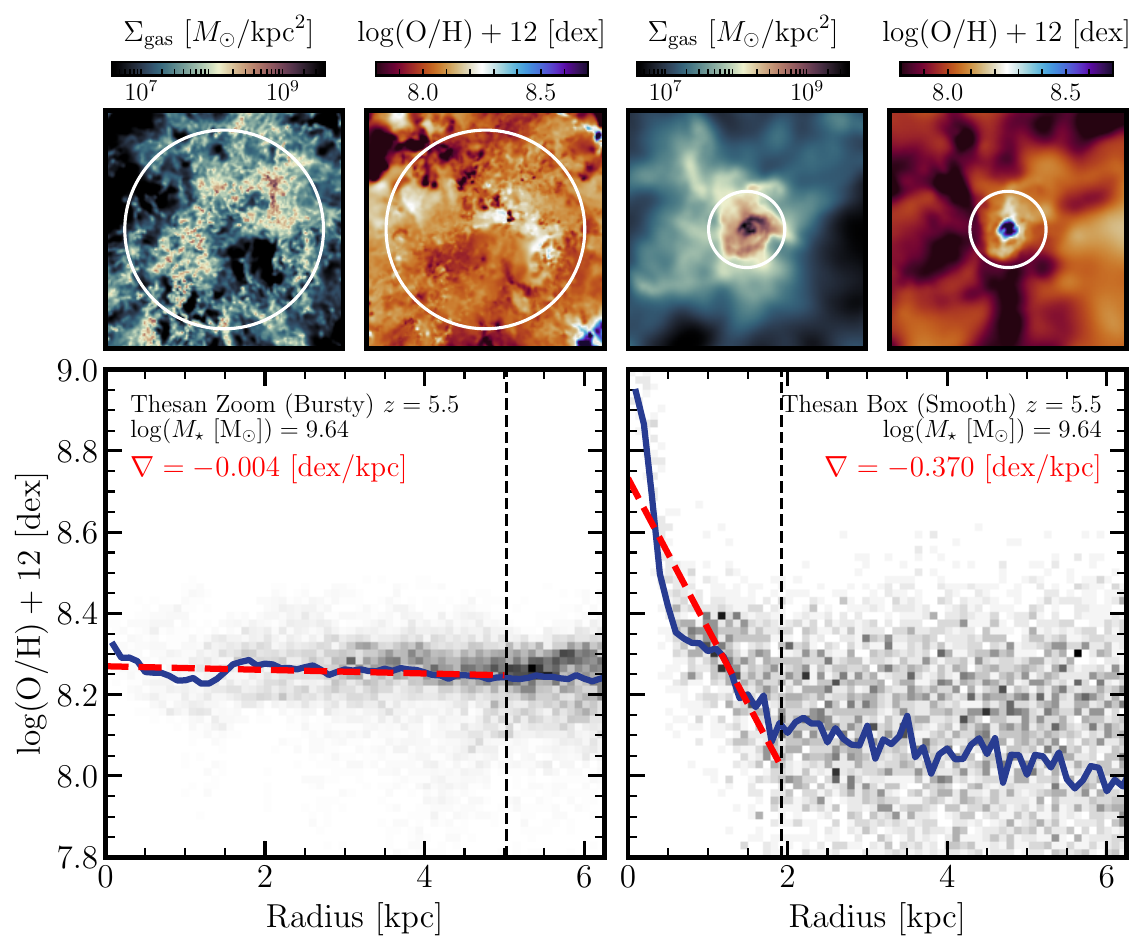}
    \caption{{\bf Example Bursty and Smooth Feedback Gradients.} 
    The gas mass surface density (top row, first and third panels) and gas-phase metallicity maps (top row second and fourth panels) within $6$ kpc for a galaxy with stellar mass $10^{9.64}~{\rm M}_\odot$ at $z=5.5$ in Thesan Zoom (left panels) and Thesan Box (right panels) simulations.
    The radial distribution of metals is shown with a 2D histogram in the background of the bottom two panels. 
    The solid blue line represents the median metallicity profile in bins of $0.1$~kpc. 
    The dashed red line shows a linear regression through the median profile at radii smaller than $R_{\rm out}$ (the radius enclosing 90\% of all star formation in the galaxy).
    The slope of the dashed line is our reported metallicity gradient.
    We quote the gradient for both galaxies in the bottom panels, finding that the Thesan Zoom galaxy has a significantly flatter gradient than that of Thesan Box. 
    }
    \label{fig:gradient_comparison}
\end{figure*}

\subsection{Simulations}
\label{subsec:simulations}

\subsubsection{FIRE}

The FIRE cosmological zoom-in simulations of galaxy formation analyzed here run using the multi-physics simulation code {\sc gizmo} (\citeauthor{Hopkins_2015} \citeyear{Hopkins_2015}) and the meshless finite mass scheme for hydrodynamics.
The FIRE-2 physics model we employ here \citep{Hopkins_2018} builds upon the original FIRE model \citep{Hopkins_2014}.

Star formation in the FIRE model occurs in dense, self-gravitating, Jeans-unstable molecular gas with $n_{\rm H} > 1000~{\rm cm}^{-3}$.
FIRE assumes star formation occurs at an efficiency of 100\% per free fall time; however, the star formation is regulated by feedback processes.
Newly formed star particles inherit their mass and metallicity from the gas in which they formed.
The stellar evolution models in FIRE come from {\sc starburst-99} v7.0 \citep{Leitherer_1999,Leitherer_2014} assuming a \cite{Kroupa_2002} IMF with masses in the range $0.1-100 {\rm M}_\odot$.
Feedback from stars is implemented through four different channels: {\it (i)} radiation pressure, {\it (ii)} supernovae with rates set by {\sc starburst-99} for Type II and \cite{Mannucci_2006} for Type Ia, {\it (iii)} stellar winds from OB and AGB stars with rates set by {\sc starburst-99}, and {\it (iv)} photo-electric heating.
The feedback events return mass and metals to the ISM as well as drive galactic winds and outflows.
The FIRE model explicitly tracks the evolution of 11 chemical species (H, He, C, N, O, Ne, Mg, Si, S, Ca, and Fe).
The yields of these species from stellar feedback events come from \cite{van_den_hoek_1997}, \cite{Marigo_2001}, and \cite{Izzard_2004} for stellar winds, \cite{Nomoto_2006} for core-collapse supernovae, and \cite{Iwamoto_1999} for Type Ia supernovae.
All of the FIRE simulations analyzed in this work implement an explicit model for unresolved turbulent metal diffusion \citep{Hopkins_2017,Su_2017,Escala_2018}.

We use data products from the FIRE-2 public data release \citep{Wetzel_2023,Wetzel_2025}, for which several galaxies were made public as part of the ``core \edit{suite}'' and ``high-redshift \edit{suite}'' (for the complete details on each \edit{suite} we refer the reader to \citeauthor{Wetzel_2023} \citeyear{Wetzel_2023}, \citeyear{Wetzel_2025}).
Our main investigation of the FIRE model in this work focuses on the high-redshift \edit{suite}: 22 galaxies run down to $z=5$\edit{,}
6 galaxies run down to $z=7$,
and 6 galaxies run down to $z=9$.
However, for Section~\ref{subsec:13Gyr} we also employ data products from 14 galaxies from the core \edit{suite}.
The simulations we utilize here come from: \citeauthor{Wetzel_2016} (\citeyear{Wetzel_2016}), \citeauthor{Garrison_Kimmel_2017} (\citeyear{Garrison_Kimmel_2017}), \citeauthor{Chan_2018} (\citeyear{Chan_2018}), \citeauthor{El_Badry_2018} (\citeyear{El_Badry_2018}), \citeauthor{Hopkins_2018} (\citeyear{Hopkins_2018}), \citeauthor{Ma_2018} (\citeyear{Ma_2018}), \citeauthor{Garrison_Kimmel_2019} (\citeyear{Garrison_Kimmel_2019}), \citeauthor{Ma_2019} (\citeyear{Ma_2019}), \citeauthor{Ma_2020} (\citeyear{Ma_2020}), and \citeauthor{Samuel_2020} (\citeyear{Samuel_2020}).
The name of the simulation corresponds to the approximate halo mass at the final snapshot ($z=0$ for the core \edit{suite} and $z=5,7,9$ for the high-redshift \edit{suite}).
The simulations were run at a number of different baryon mass resolutions, ranging from $\sim100~{\rm M}_\odot$ for the smaller halos to $\sim7000~{\rm M}_\odot$ for larger halos.
We refer the reader to \citeauthor{Wetzel_2023} (\citeyear{Wetzel_2023}) and the works listed above for more details on each simulation model.
Finally, we note that the FIRE-2 simulations we use here do not include AGN feedback, magnetic fields, or cosmic rays.


For the analysis of FIRE, we use both the \textsc{gizmo analysis} and \textsc{halo analysis} tools provided by \cite{Wetzel_2016}.

\subsubsection{SPICE}

We also use data products from the SPICE cosmological simulations \citep{Bhagwat_2024}.
SPICE is a series of cosmological box simulations run with the {\sc ramses-rt} code \citep{Rosdahl_2013, Rosdahl_2015}, which is an extension of the original {\sc ramses} adaptive mesh refinement code \citep{Teyssier_2002}.
The main advantages of the SPICE simulations \edit{are} their variations in the treatment of stellar feedback.
SPICE implements several different feedback models, two of which we investigate here: smooth-sn and bursty-sn (hereafter ``SPICE Smooth'' and ``SPICE bursty'', respectively).
The SPICE Smooth model allows for stars within the particles to explode anywhere from 3--40 Myr after birth (depending on their mass and rates set by {\sc starburst-99}), leading to smoother injection of energy back into the ISM.
The SPICE Bursty model, on the other hand, assumes that all supernovae go off in a single event 10 Myr after birth.
Populations of stars exploding concurrently make stellar feedback particularly efficient in the SPICE Bursty model.
It should be noted that the SPICE approach of modifying delay time distributions of supernovae is different than the bursty feedback in FIRE and Thesan Zoom.
However, the key advantage of SPICE for the purposes of this work is the same physical model with different feedback implementations.

Stars form in the SPICE model according to prescriptions set in \cite{Kretschmer_2020} which incorporate a model for (unresolved) turbulence \citep{Schmidt_2006} allowing for variable star formation rate efficiencies that depend on the local conditions of the ISM (i.e., the virial parameter and turbulent mach number; \citeauthor{Hennebelle_2011} \citeyear{Hennebelle_2011}, \citeauthor{Federrath_2012} \citeyear{Federrath_2012}).
Star formation in SPICE therefore occurs in two modes: in gravitationally unstable gas and regions with large amounts of supersonic turbulence.
This star formation occurs in dense ($n_{\rm H}\geq10~{\rm cm}^{-3}$) gas and forms stars according to a \cite{Chabrier_2003} IMF.
Feedback from these stars comes in the form of radiative energy (photo-ionization and -heating) and radiative pressure in addition to the aforementioned supernova feedback variations.
Metals are advected across cell boundaries as passive scalars \citep[following from][]{Rosdahl_2013}.
Finally, we note that SPICE does not implement feedback from black holes.
Each of the SPICE boxes we analyze here have a volume of $(10~{\rm cMpc}/h)^3$ at initial baryon mass resolution of $975\,{\rm M}_\odot$.

\subsubsection{Thesan}
\label{subsubsec:thesan}

Finally, we use data products from the Thesan suite of simulations.
Thesan consists of a large $(95.5~{\rm cMpc})^3$ box \citep[][hereafter Thesan Box]{Garaldi_2022,Kannan_2022,Smith_2022}, as well as a series of high resolution zoom-in simulations \citep[][hereafter Thesan Zoom]{Kannan_2025}.
Both the box and zooms are run using the radiation hydrodynamics moving-mesh code {\sc arepo-rt} (\citeauthor{Kannan_2019} \citeyear{Kannan_2019}; based on the original {\sc arepo} code \citeauthor{Springel_2010} \citeyear{Springel_2010}).
Both Thesan Box and Thesan Zoom add an additional model for the creation and destruction of dust from gas according to \cite{McKinnon_2016,McKinnon_2017}.
While the details of the dust module are beyond the scope of this work, it is important to note that the dust forms out of the metals in the gas phase.
Therefore, a (small) component of the metals are lost to dust formation, which we take to be negligible here.
Moreover, both Thesan Box and Thesan Zoom explicitly track the evolution of several chemical species (H, He, C, N, O, Ne, Mg, Si, and Fe) with yields coming from \cite{Nomoto_1997} for Type Ia supernovae, \cite{Portinari_1998} and \cite{Kobayashi_2006} for Type II supernovae, \cite{Karakas_2010}, \cite{Doherty_2014}, and \cite{Fishlock_2014} for AGB winds.

\subsubsection*{\texorpdfstring{ T\lowercase{hesan} B\lowercase{ox}}{Thesan Zoom}}
Thesan box is run using a modified version of the IllustrisTNG galaxy formation model \citep{Pillepich_2018a}, which incorporates the physics of both stellar and AGN feedback.
Stars form in the dense ($n_{\rm H} \gtrsim 0.1~{\rm cm}^{-3}$) ISM according to the \cite{Springel_Hernquist_2003} equation of state and a \cite{Chabrier_2003} IMF.
The \cite{Springel_Hernquist_2003} equation of state provides pressure support for the ISM and sets the efficiency at which stars form from the dense gas following an empirical Kennicutt-Schmidt relation \citep{Schmidt_1959,Kennicutt_1998}.
Feedback from these stars is implemented in the form of supernovae explosions and stellar winds in the form of kinetic and thermal energy \citep[see][]{Pillepich_2018a}.
Black holes with mass $1.18\times10^6~{\rm M}_\odot$ are seeded in halos exceeding $7.38\times10^{10}~{\rm M}_\odot$ in the Thesan Box model \citep{Weinberger_2018}.
Feedback from black holes is modeled via two channels based on the accretion rates of the black holes: the high accretion thermal mode and low accretion kinetic mode \citep[see][]{Weinberger_2018}.
The thermal mode continuously dumps thermal energy into the ISM and dominates for low mass systems, whereas high mass galaxies are dominated by the directed and pulsed kinetic winds.
We utilize the highest resolution Thesan-1 volume which is a $(95.5~{\rm cMpc})^3$ box with $2\times2100^3$ particles, corresponding to an initial baryon mass resolution of $3.12\times10^6~{\rm M}_\odot$.

\subsubsection*{\texorpdfstring{ T\lowercase{hesan} Z\lowercase{oom}}{Thesan Zoom}}

Thesan Zoom employs a tailored version of the SMUGGLE galaxy formation model \citep{Marinacci_2019}, which shares a number of similarities with FIRE.
Stars form in dense ($n_{\rm H} > 10~{\rm cm}^{-3}$), self-gravitating, Jeans-unstable gas with an efficiency per free-fall time of 100\%.
Similar to FIRE, the global star formation is self-regulated by feedback from the newly formed stars and maintains a lower efficiency compared to the prescribed local value \citep{Shen_2025,Wang_2025}.
Stellar feedback in Thesan Zoom is modeled through a number of different channels: radiative feedback \citep{Kannan_2020}, stellar winds and supernovae \citep{Marinacci_2019}, and ``early'' stellar feedback in the form of momentum driven winds (\citeauthor{Kannan_2025} \citeyear{Kannan_2025}).
Notably, the Thesan Zoom suite includes several variations in these feedback prescriptions; however, for this work, we use only the fiducial model.
We note that Thesan Zoom does not include contributions from AGN.
The Thesan Zoom Suite consists of 14 halos selected from the Thesan Box run down to $z\sim3$ at various baryon mass resolutions ranging from $142~{\rm M}_\odot$ to $9.09\times10^3~{\rm M}_\odot$ (see \citeauthor{Kannan_2025} \citeyear{Kannan_2025} their Table 2).
We make use of the highest resolution versions of each halo available (see \citeauthor{Kannan_2025} \citeyear{Kannan_2025} their Table 3).

\subsection{Galaxy Selection Criteria}
\label{subsec:selection}

The three simulation models of this work adopt the following halo finding algorithms: SPICE uses AdaptaHOP, FIRE adopts Amiga Halo Finder \citep{Knollmann_2009}, and Thesan employs Subfind \citeauthor{Springel_2001} (\citeyear{Springel_2001}; Thesan Zoom uses the updated Subfind-hbt \citeauthor{Springel_2021} \citeyear{Springel_2021}).
While the structure finding algorithms are not the same and can thus assign different particles/cells to the various (sub)structures \citep[e.g.,][]{Forouhar_2025}, we do not anticipate the detailed properties of the very inner regions of galaxies ($\sim$few kpc) to be significantly impacted by the different structure finders.

The simulations also have different resolutions, even within the same galaxy formation model (e.g., Thesan Zoom and FIRE).
To ensure that our galaxies are well-resolved, we require a threshold of $>10^3$ gas and star particles within the system \citep[following from][]{Garcia_2025b}.
Roughly speaking, this corresponds to minimum stellar and gas masses of $\sim 10^{6}~{\rm M}_\odot$ in SPICE as well as high resolution FIRE and Thesan Zoom runs, $\sim 10^{7}~{\rm M}_\odot$ in lower resolution FIRE and Thesan Zoom runs, and $\sim10^{8}~{\rm M}_\odot$ in the Thesan Box.
Moreover, since emission line metallicity diagnostics come from star-forming {\sc Hii} regions \citep[e.g.,][]{Kewley_2019}, we require that our galaxies have instantaneous non-zero star formation rates.
Finally, we note that we restrict our analysis to the most massive subhalo within each group (i.e., the central galaxy).


\subsection{Gradient Definition}
\label{subsec:gradient_definitions}

We largely follow the methodology of \citeauthor{Garcia_2025b} (\citeyear{Garcia_2025b}; which itself derives heavily from a combination of \citeauthor{Ma_2017} \citeyear{Ma_2017} and \citeauthor{Hemler_2021} \citeyear{Hemler_2021}) to define the metallicity gradients with only a few minor changes.
We first briefly describe the methodology employed in this work, then note any changes from the previous methodology, and briefly discuss how they may impact our results.
We demonstrate the application of these methodologies on two galaxies in Figure~\ref{fig:gradient_comparison}.

We first center the galaxy \edit{on the location of the potential minimum} and rotate to a face-on orientation.
To rotate the galaxy to face-on, we define the angular momentum vector of star-forming gas within a region $R_{\rm in} < r < R_{\rm out}$ (where $R_{\rm in}$ and $R_{\rm out}$ are the regions containing 5\% and 90\% of the total star formation of the galaxy \edit{within 10 kpc}, respectively).
We then align this angular momentum vector along the $+z$ axis.
We note that the concept of ``face-on'' for galaxies with bursty feedback is not as well posed as for the more disk-like galaxies of smooth feedback models (see, e.g., Figure~\ref{fig:gradient_comparison}).
To make as fair a comparison as possible, we re-orient all galaxies in each model according to the above prescription.
Since the ISM of bursty feedback galaxies lacks coherent structure, we do not anticipate that the (re-)orientation plays a significant role in our results.

With our face-on galaxy, we create two-dimensional mass-weighted metallicity maps and gas mass maps using pixel sizes of $0.1~{\rm kpc}\times0.1~{\rm kpc}$ (shown in the top row of Figure~\ref{fig:gradient_comparison}).
We then convert the two-dimensional maps into one-dimensional radial distributions~(gray background distribution in large panels of Figure~\ref{fig:gradient_comparison}).
We further reduce the radial distribution of metals into a singular median metallicity profile in bins of $0.1$ kpc~(solid blue lines in large panels of Figure~\ref{fig:gradient_comparison}).
We note that, since metallicity emission line diagnostics come from star-forming regions in galaxies, we restrict our sample to pixels with gas mass surface densities $>10^6~{\rm M}_\odot\,{\rm kpc}^{-2}$ \citep[which are likely to host star-forming gas;][]{Ma_2017}.
Finally, we summarize the median metallicity profile by fitting a linear regression (in logarithmic metallicity) in the region $r < R_{\rm out}$~(dashed red lines in Figure~\ref{fig:gradient_comparison}).
We note that a single linear regression is a simplistic summary statistic for metallicity profiles~(see, e.g., bottom right panel of Figure~\ref{fig:gradient_comparison}).
Many of the profiles flatten out at large radii \citep[see also][]{Garcia_2023} and/or may have multiple components \citep[see also][]{Tapia_2025}; however, a detailed examination of the shapes of the profiles of these simulations is beyond the scope of the current work.

The largest difference between the methodology applied in this work and that of \cite{Garcia_2025b} is that here we do not truncate the gradient measurement at an inner radius.
The reasoning behind ignoring the central regions of galaxies in the previous work was primarily to eliminate the impact of AGN, which can significantly suppress metals in the inner regions and contaminate emission line diagnostics. This practice is commonly adopted in low redshift studies \citep[e.g.,][]{Sanchez_2012,Sanchez_2014,Sanchez_Menguiano_2016,Belfiore_2017}.
At higher redshift, however, it is less common to exclude the inner regions \citep[e.g.,][]{Vallini_2024,Venturi_2024,Fujimoto_2025} and most of the simulations here do not model AGN (although Thesan Box does).\ignorespaces
\footnote{\ignorespaces
While Thesan Box does model AGN feedback, we do not expect that it will have a significant impact on the galaxies in our sample.
As mentioned in Section~\ref{subsubsec:thesan}, the kinetic mode AGN feedback \citep[which is the mode most dominant in quenching galaxies and driving the interior metallicity variations;][]{Weinberger_2017} effectively only impacts massive galaxies, thus the impact of AGN should be minimal in our sample.
}
\edit{We therefore relax the inner region criterion for this work, finding that the inclusion/exclusion of the inner radial cut does not significantly impact our core results.} 
Finally, we note that SPICE does not trace the evolution of individual metal species.
The gradients we quote in SPICE are therefore the {\it total} metallicity gradient scaled by assuming a $35\%$ fraction of oxygen in metals and a total mass fraction that is $76\%$ hydrogen as opposed to the direct oxygen-to-hydrogen abundances of the other models.
In practice, there is generally only a modest $(\lesssim0.1~{\rm dex)}$ difference between the the scaled total metallicity and direct oxygen-to-hydrogen abundances \citep[see][Appendix B and Figure B2]{Garcia_2025a}.


\section{Results}
\label{sec:results}

\subsection{Distribution of Metallicity Gradients in each simulation}
\label{subsec:distribution_gradients}

\begin{figure*}
    \centering
    \includegraphics[width=\linewidth]{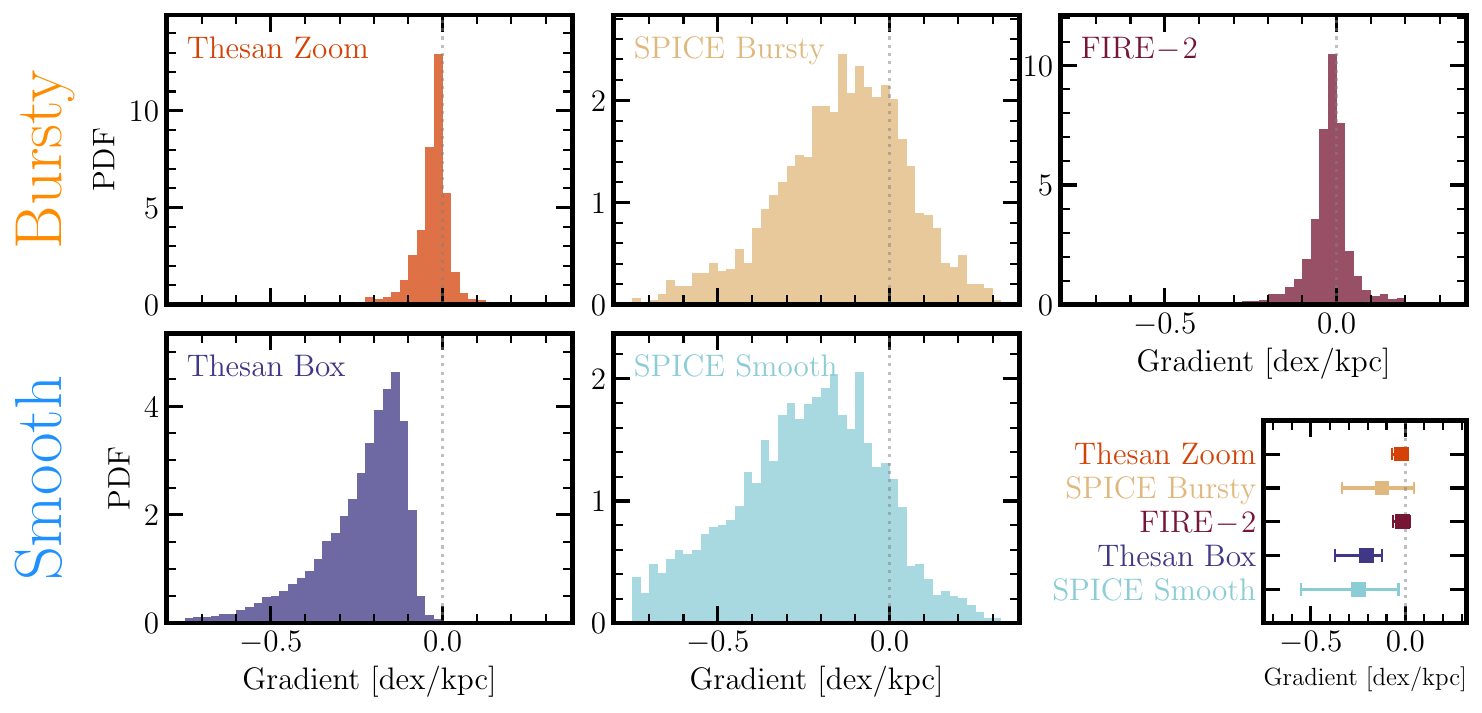}
    \caption{{\bf Distribution of Gas-Phase Radial Metallicity Gradients in Bursty and Smooth Feedback Models at High-Redshift ($3<z\lesssim 11$).}
    Our sample of metallicity gradients in Thesan Zoom (top left), SPICE Bursty (top middle), FIRE-2 (top right), Thesan Box (bottom left), and SPICE Smooth (bottom middle).
    The top row represent the bursty feedback models, while the bottom row represents the smooth feedback models.
    We summarize the distributions in the small figure in the bottom right with the median, $16^{\rm th}$ percentile, and $84^{\rm th}$ percentile of the distributions.
    We find that the gradients in the smooth feedback models are systematically more negative than their bursty feedback counterparts.
    }
    \label{fig:all_gradients}
\end{figure*}

We begin by characterizing the  metallicity gradients in each simulation.
Figure~\ref{fig:all_gradients} shows the probability density function (PDF) of all of the measured gradients in the bursty feedback models (top left is Thesan Zoom, top middle is SPICE Bursty, and top right is FIRE) and the smooth feedback models (bottom left is Thesan Box and bottom middle is SPICE Smooth).
We note that combining the gradients in this way is a crude comparison, since it ignores several important factors, e.g., the redshift of the galaxy or its stellar mass (both of which we explore in more detail in Section~\ref{subsec:stellar_mass}).
Despite the crudeness of the comparison, we find a qualitative difference in the gradients measured in the smooth feedback models to those of the bursty feedback models, with the latter typically having flatter gradients.

To quantify the discrepancy between the two different types of simulations, the bottom right panel of Figure~\ref{fig:all_gradients} shows the median and spread (taken to be the $16^{\rm th}$ and $84^{\rm th}$ percentiles of the distributions) for each model analysed in this work.
We find that the median gradients are $-0.021^{0.008}_{-0.072}$ dex/kpc in Thesan Zoom, $-0.124_{-0.333}^{0.048}$ dex/kpc in SPICE Bursty, 
\edit{$-0.016_{-0.066}^{0.022}$}
dex/kpc in FIRE, $-0.205_{-0.373}^{-0.124}$ dex/kpc in Thesan Box, and $-0.248_{-0.553}^{-0.037}$ dex/kpc in SPICE Smooth.
On the whole, the bursty feedback models produce significantly flatter than their smooth counterparts: for the Thesan simulations, the gradients in the Box are approximately an order of magnitude steeper than their Zoom counterparts, while the SPICE Smooth gradients are steeper than the SPICE Bursty by a factor of $\sim2$.
Moreover, the width of the distribution of gradients in the bursty models is significantly smaller than those of the smooth models, with the exception of SPICE which have comparable scatter (we investigate the discrepancy of the SPICE results further in Section~\ref{subsec:stellar_mass}).
In each of the samples, we find that the distributions tend to be skewed towards more negative gradients, with the $16^{\rm th}$ percentiles being further from the median than the $84^{\rm th}$ percentiles.
Clearly, even without more careful considerations for the detailed properties of the individual systems, there is a qualitative difference between the metallicity gradients from bursty and smooth feedback models.
The bursts of feedback from stars seem to efficient redistributing metals through the galaxy, whereas the smooth feedback models do not.

It is also noteworthy that the gradients are remarkably consistent across similar models, regardless of the details of the simulation implementation.
This similarity echoes the low redshift, large box simulations we investigated previously \citep{Garcia_2025b}.
While each simulation model---both in this work and in \cite{Garcia_2025b}---is attempting to model the same general physics (e.g., star formation, stellar feedback, chemical enrichment), there is a wide diversity of numerical implementations and assumptions that go into the different models (see, e.g., Section~\ref{subsec:simulations}).
Yet, despite these differences, the metallicity gradients in each of the models have highly self-similar trends: smooth feedback models have strong negative gradients while bursty feedback models have flatter gradients (this similarity is also shared with the lower redshift simulation volumes, which we discuss in more detail in Section~\ref{subsec:13Gyr}).
There is, of course, variation between the predictions when considering properties of the host galaxies (which we discuss more in the next section); however, the level of out-of-the-box agreement is worth appreciating.
The agreement is especially remarkable given that the total metal budget of galaxies in different simulation models can vary substantially \citep{Garcia_2024b,Garcia_2025a}.
Thus, we emphasize the utility of metallicity gradients as a diagnostic for stellar feedback at high-redshift ($z\gtrsim3$), which we discuss in more detail in Section~\ref{subsec:13Gyr}.

\subsection{Gradients as a Function of Stellar Mass and Redshift}
\label{subsec:stellar_mass}

\begin{figure*}
    \centering
    \includegraphics[width=\linewidth]{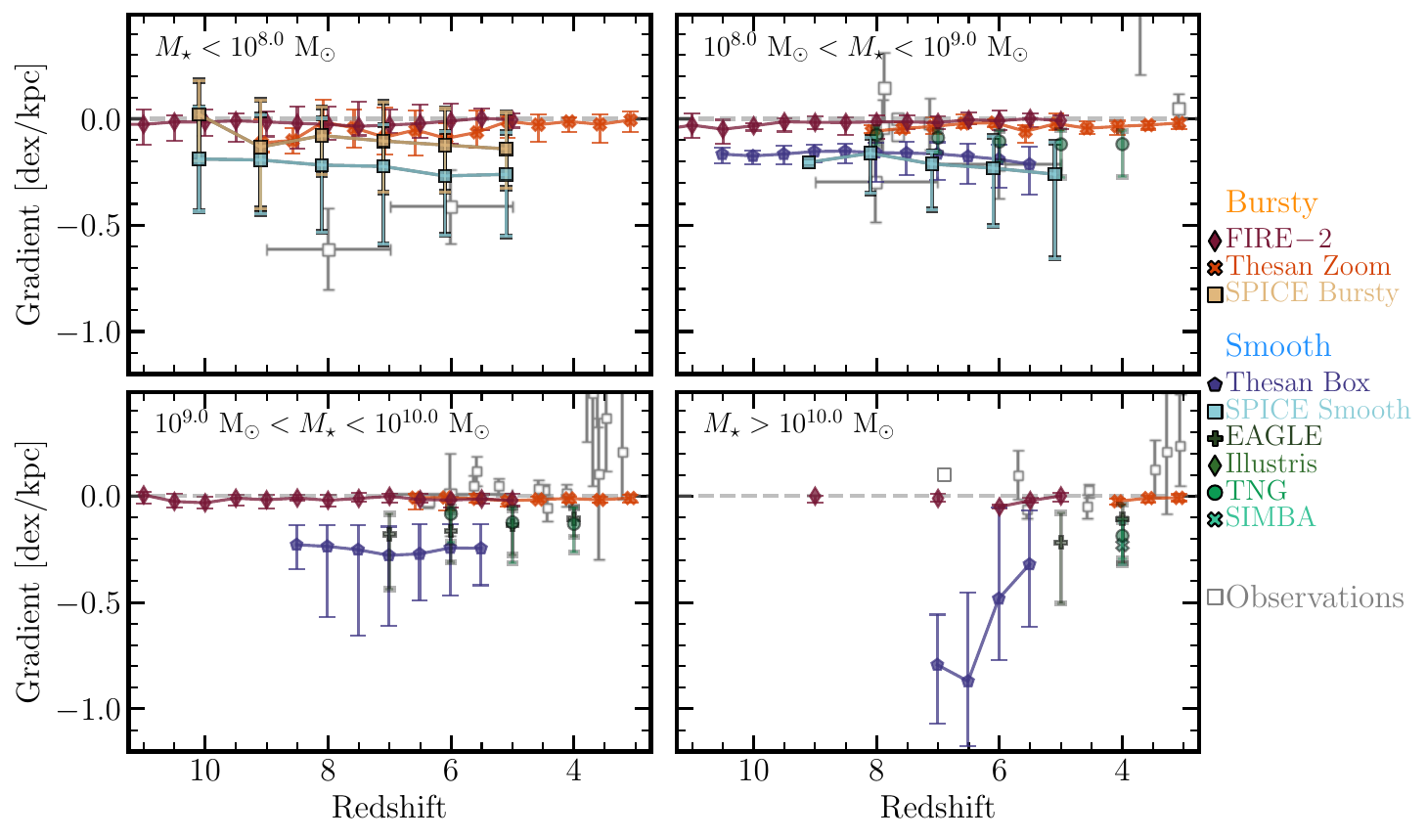}
    \caption{{\bf Metallicity Gradient Evolution Broken into Four Stellar Mass Bins.} 
    The gradient evolution in FIRE-2 (triangles), SPICE (squares), and Thesan (solid lines) broken into stellar mass bins: $M_\star < 10^{8}~{\rm M}_\odot$ (top left), $10^{8}~{\rm M}_\odot < M_\star < 10^{9}~{\rm M}_\odot$ (top right), $10^{9}~{\rm M}_\odot < M_\star < 10^{10}~{\rm M}_\odot$ (bottom left), and $M_\star > 10^{10}~{\rm M}_\odot$ (bottom right).
    We also include a recent series of observed metallicity gradients from {\it JWST} \protect\citep[when stellar mass estimates available;][]{Troncoso_2014,Arribas_2024,Venturi_2024,Acharyya_2025,Fujimoto_2025,Ivey_2025,Li_2025} and the relations from EAGLE, Illustris, TNG, and SIMBA at $z>3$ from \protect\cite{Garcia_2025b}.
    }
    \label{fig:stellar_mass}
\end{figure*}

We now more carefully consider our sample of galaxies.
The main properties that we investigate here are: {\it (i)} the stellar mass of the host galaxy, and {\it (ii)} the redshift of the galaxy.
Figure~\ref{fig:stellar_mass} shows the metallicity gradients of galaxies in each of the simulation models as a function of redshift, broken into four different stellar mass bins corresponding to galaxies in that mass bin at the given redshift (i.e., not explicitly tracking progenitors through time).
We only plot values at integer and half-integer redshifts (e.g., $z=5.0$ and $z=5.5$) for clarity, but data from Thesan Box, Thesan Zoom, and FIRE includes gradients measured at intermediate redshifts (e.g., $z=5.2$).
Finally, we also note that we require more than five galaxies in each bin to plot it.

The gradients in SPICE Smooth model in the lowest mass bin ($M_\star < 10^8~{\rm M}_\odot$; top left of Figure~\ref{fig:stellar_mass}) have virtually no dependence on redshift, with an average value of $\sim-0.2$ to $-0.3$~dex/kpc.
The spread in gradients in this stellar mass bin for SPICE Smooth is significant, covering $\pm0.3$~dex/kpc.
The galaxies in SPICE Smooth therefore cover a wide diversity of positive, negative, and flat gradients in this mass bin.
The bursty feedback models---FIRE, SPICE Bursty, and Thesan Zoom---also have very little redshift evolution in this mass bin and have flatter gradients on average than SPICE Smooth.
The scatter about these gradients is less than SPICE Smooth, covering approximately $\pm0.1$~dex/kpc in FIRE and Thesan Box, whereas the scatter in SPICE Bursty is comparable to SPICE Smooth.
On the whole, we find that there are no significant differences between the metallicity gradients of smooth and bursty feedback models in this lowest mass bin.

In the next mass bin ($10^{8}~{\rm M}_\odot < M_\star < 10^9~{\rm M}_\odot$; top right of Figure~\ref{fig:stellar_mass}), the smooth feedback models have qualitatively similar average gradients as the lowest mass bin: around $-0.2$~to~$-0.3$~dex/kpc with little redshift evolution.
The distribution of gradients $(\pm 0.1~{\rm dex/kpc})$, though, is much smaller than in the lower mass bin at $z\gtrsim7$, but is comparable to the previous mass bin at $z=5-6$.
In the bursty feedback models, we also find qualitatively similar behavior as the lowest mass bin: flat gradients with very little redshift evolution.
The full distribution of gradients in the bursty feedback models is approximately the same as the lowest mass bin $\sim \pm0.1$~dex/kpc.
In this mass bin, there is more of a difference between the smooth and bursty models, but the difference is subtle.

The difference between the smooth and bursty models is more apparent in the next mass bin ($10^9~{\rm M}_\odot < M_\star < 10^{10}~{\rm M}_\odot$; bottom left of Figure~\ref{fig:stellar_mass}).
The smooth feedback models still have virtually no redshift evolution in this mass bin and their average gradients are of the order of $-0.3$~dex/kpc, but the distribution of gradients increases at lower redshift (possibly due to the increased sample size).
Specifically, there exist many more very strong $\sim -0.5$~dex/kpc gradients and virtually no gradients shallower than $-0.15$~dex/kpc.
The distribution of gradients changes for the bursty feedback models, too, compared to the lowest mass bin.
The width of the distribution shrinks to $\sim0.05$~dex/kpc.
Here, the {\it qualitative} difference -- opposed to quantitative difference in the lowest mass bin -- between the smooth and bursty feedback models becomes more apparent.
There are very few flat gradients in the smooth feedback models in this stellar mass range (only $\sim2\%$ of galaxies have gradients shallower than $-0.075$~dex/kpc) and there are virtually no strong negative gradients in the bursy feedback models (only $\sim3\%$ of galaxies have gradients steeper than $-0.075$~dex/kpc).

Finally, the difference between the smooth and bursty feedback models is most pronounced in the highest mass bin ($M_\star > 10^{10}~{\rm M}_\odot$; bottom right of Figure~\ref{fig:stellar_mass}).
At this stellar mass there are only a few galaxies from the bursty sample\ignorespaces\footnote{
The lack of high stellar mass galaxies in the bursty models can be attributed to: {\it (i)} that bursty feedback models produce galaxies with lower stellar mass at fixed halo mass (\citeauthor{Bhagwat_2024} \citeyear{Bhagwat_2024}) and {\it (ii)} there are only a few massive galaxies in the -- already small -- FIRE-2 and Thesan Zoom samples.
}
Regardless, the few galaxies in FIRE and Thesan Zoom still have very flat gradients with very little scatter.
For the smooth feedback models, we find a strong relationship between gradient and redshift in this most massive bin.
Galaxies in Thesan Box have (already quite strong) gradients of $\sim-0.4$~dex/kpc at $z=5-5.5$, which increases to average gradients of $\sim-0.8$~dex/kpc at $z=7$.
At $z=7$, there is a difference of approximately an order of magnitude in the metallicity gradients of the galaxies.
We investigate this in more detail, and compare with the one observed galaxy at this redshift, in Section~\ref{subsec:comparison_observations}.

In all, we find that the level to which the metallicity gradients of galaxies in smooth and bursty feedback models disagree depends on stellar mass.
We find that the gradients in galaxies with the largest stellar masses ($M_\star \gtrsim 10^9~{\rm M}_\odot$) are the most discordant between the different models.
Low mass galaxies ($M_\star \lesssim10^9~{\rm M}_\odot$), on the other hand, are only quantitatively different, not qualitatively, with the very low mass galaxies ($M_\star < 10^8~{\rm M}_\odot$) being largely indistinguishable between the different feedback types.

\section{Discussion}
\label{sec:discussion}

\subsection{13 Billion Years of Gradient Evolution from Cosmological Simulations}
\label{subsec:13Gyr}

\begin{figure*}
    \centering
    \includegraphics[width=0.95\linewidth]{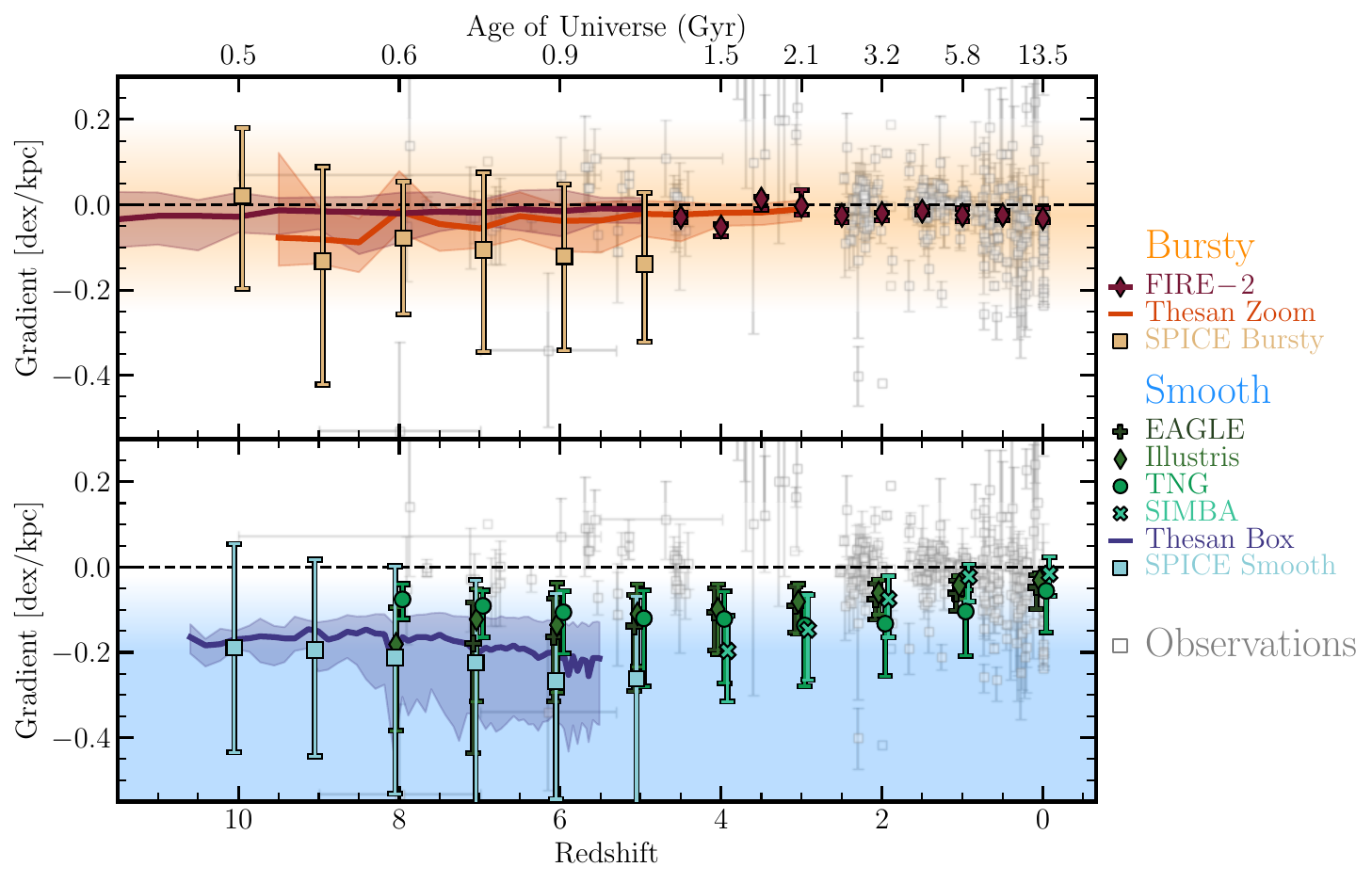}
    \caption{{\bf 13 Billion Years of Gas-Phase Metallicity Gradient Evolution in Cosmological Simulations.}
    The composite evolution of metallicity gradients from this work and \protect\cite{Garcia_2025b} in ({\bf Top}) bursty feedback models FIRE-2 (maroon diamonds), Thesan Zoom (solid orange line), and SPICE Bursty (orange squares), as well as ({\bf Bottom}) smooth feedback models EAGLE (plus), Illustris (green diamonds), TNG (circles), SIMBA (Xs), Thesan Box (solid blue line), and SPICE Smooth (blue squares).
    The lines/markers represent the median gradient at a particular redshift while the shaded regions/error bars represent the width of the $16-84^{\rm th}$ percentiles.
    In general, we find that bursty feedback models tend to have flatter metallicity gradients than their smooth feedback model companions, regardless of the detailed implementation of the models.
    The shaded regions in each panel {\it qualitatively} show where the majority of each feedback type's gradients lie, particularly at high-redshift.
    We also present a compilation of observed metallicity gradients from \protect\cite{Rupke_2010b,Queyrel_2012,Swinbank_2012,Jones_2013,Jones_2015,Troncoso_2014,Leethochawalit_2016,Wang_2017,Wang_2019,Wang_2022,Carton_2018,Forster_Schreiber_2018,Curti_2020,Grasha_2022,Li_2022,Li_2025,Arribas_2024,Tripodi_2024,Vallini_2024,Venturi_2024,Acharyya_2025,Fujimoto_2025,Ivey_2025,Ju_2024,Vale_2025}.
    Recent high-redshift observations, with the exception of those from \protect\cite{Li_2025}, seem to favor more bursty feedback scenarios.
    }
    \label{fig:13Gyr}
\end{figure*}

Figure~\ref{fig:13Gyr} shows the redshift evolution of metallicity gradients in nine modern cosmological simulation models.
We categorize each simulation model as having either bursty or smooth stellar feedback.
The bursty feedback models are FIRE-2 \citep{Hopkins_2018}, Thesan Zoom \citep{Kannan_2025}, and SPICE Bursty \citep{Bhagwat_2024}, while the smooth feedback models are EAGLE \citep{Schaye_2015}, Illustris \citep{Vogelsberger_2014a}, IllustrisTNG \citep{Pillepich_2018a}, SIMBA \citep{Dave_2019}, Thesan Box \citep{Garaldi_2024}, and SPICE Smooth \citep{Bhagwat_2024}.
The data from the EAGLE, Illustris, TNG, and SIMBA simulations come from \citeauthor{Garcia_2025b} (\citeyear{Garcia_2025b}; see caveat about slight change in methodology in Section~\ref{subsec:gradient_definitions}).
We note that this plot also includes FIRE-2 galaxies from the ``core \edit{suite}''.
We treat these systems the same as their high-redshift counterparts and only report gradients at integer and half-integer redshifts at $z<5$.
Briefly, we find that our analysis of the FIRE-2 core \edit{suite} at $z < 5$ is consistent with previous FIRE papers, finding a diversity of metallicity gradients with a preference for flat gradients \citep{Ma_2017,Bellardini_2021,Graf_2024,Sun_2024}.

Figure~\ref{fig:13Gyr} is essentially an updated version of \citeauthor{Gibson_2013} (\citeyear{Gibson_2013}; their Figure 1), which included the evolution of four galaxies to $z\sim2.5$.
\cite{Gibson_2013} find that the systems with `enhanced' feedback have flat gradients, while systems with `conservative' feedback have strong negative gradients that get stronger further back in time.
Remarkably, we find qualitatively the same trend with the benefit of several additional simulation models with updated physics models and several orders of magnitude more galaxies.
Galaxies with bursty feedback tend to have weaker metallicity gradients than their smooth feedback counterparts, especially at high-redshift ($z\gtrsim2$) and with large stellar masses (Figure~\ref{fig:stellar_mass} and \citeauthor{Garcia_2025b} \citeyear{Garcia_2025b}, Figure 3).
The rate at which we find gradients to steepen with redshift in the smooth feedback models is approximately $-0.02~{\rm dex/kpc/}\Delta z$ \citep[in good agreement with the values of][]{Hemler_2021,Garcia_2025b}, while the bursty feedback models are roughly consistent with $0.00~{\rm dex/kpc}/\Delta z$.

We note that the strength of gradients appears to ``taper'' at the highest redshifts ($z\gtrsim8$) in the smooth feedback models, instead of continuing the trend of stronger gradients with increasing redshift.
We argue that this is likely a selection effect of the stellar masses within the samples at each redshift.
As shown in Figure~\ref{fig:stellar_mass} (and Figure~3 of \citeauthor{Garcia_2025b} \citeyear{Garcia_2025b}), lower mass galaxies $(M_\star \lesssim 10^9~{\rm M}_\odot)$ in smooth feedback models tend to have flatter gradients than their high-mass counterparts (although it is unclear if this holds for galaxies at all redshifts for very massive $M_\star > 10^{11}~{\rm M}_\odot$ galaxies; see \citeauthor{Garcia_2025b} \citeyear{Garcia_2025b}).
The sample of galaxies in, e.g., Thesan Box and SPICE Smooth tends towards lower mass systems at these very early times.
It is worth noting that this same effect was noticed previously within the TNG suite, where the composite evolution of the gradients depended more on the mass composition of the sample than, e.g., resolution considerations \citep[see discussion in Appendix A and Figure 5 of][]{Garcia_2025b}.
We therefore suggest that, given a sufficiently large volume, it is likely that the gradients would continue to steepen at redshifts $z>8$ and that the observed tapering is just a selection effect from the cosmological volumes.

\subsubsection{Subtle Differences Between Bursty Feedback Models}
\label{subsubsec:bursty_model_comparisons}

\edit{
While we find that there is a qualitatively different character to metallicity gradients between the bursty and smooth feedback models, there are still differences within each classification.
Here, we briefly comment on a potential difference in the bursty feedback models leading to the detailed differences discussed in Section~\ref{subsec:distribution_gradients} (noting that a more in depth discussion of the smooth feedback models can be found in; \citeauthor{Garcia_2025b} \citeyear{Garcia_2025b}).
}

\edit{
As mentioned previously, the SPICE Bursty gradients tend to, on average, have more negative gradients than either Thesan Zoom or FIRE (see, e.g., Figures~\ref{fig:all_gradients}~and~\ref{fig:stellar_mass}).
In particular, the $16^{\rm th}$ percentile of the distribution in SPICE Bursty is $-0.333~{\rm dex/kpc}$, which is comparable to the median of both the Thesan Box and SPICE Smooth gradients, whereas Thesan Zoom and FIRE have very few systems wtih gradients more negative than $-0.1~{\rm dex/kpc}$.
Moreover, the distribution of gradients in Thesan Zoom and FIRE-2 is significantly narrower than in SPICE Bursty ($\sim0.08~{\rm dex/kpc}$ in Thesan Zoom, $\sim0.09~{\rm dex/kpc}$ in FIRE, and $\sim0.37~{\rm dex/kpc}$ in SPICE Bursty).
}

\edit{
It is difficult to pinpoint exactly the cause(s) of differences in gradients between the simulations as the distribution of metals in a galaxy is highly sensitive to the underlying physical processes.
However, one likely candidate is how turbulence is handled in these models.
SPICE implements a subgrid model for turbulence \citep[that of][]{Schmidt_2006} which tracks the evolution of unresolved turbulent kinetic energy and couples it to the local gas thermodynamics, thereby modeling the impact of small-scale turbulent pressure and mixing below the grid scale.
This introduces an additional turbulent pressure component and an effective mixing scale that depend on the local turbulent velocity and dissipation timescale which are evaluated at the grid scale, limiting metal transport to neighboring cells.
In contrast, the FIRE and Thesan Zoom models capture turbulence through resolved gas dynamics, with stellar feedback small scale turbulence as well as bulk flows in the gas, rather than having a subgrid turbulence model.
These feedback-driven winds can drive large-scale outflows, fountain cycling, and sustained turbulence \citep[e.g.,][]{Muratov_2015} and  redistribute metals over larger spatial scales following burst episodes \citep{Ma_2017,Muratov_2017}.
It is therefore possible that the comparatively tight metallicity-gradient distributions in Thesan Zoom and FIRE reflect more efficient large-scale radial mixing and recycling than in SPICE Bursty.
In this light, the simulation models analyzed in this work and \cite{Garcia_2025b} fall into one of three categories: (i) do not have subgrid turbulence models (Illustris, IllustrisTNG, EAGLE, SIMBA, Thesan Box), (ii) have a subgrid turbulence model (SPICE), and (iii) have explicit feedback-driven turbulence (FIRE and Thesan Zoom).
In addition to the qualitative, bursty versus smooth, nature of feedback, metallicity gradients likely also present a unique laboratory for understand the small-scale turbulence present in the ISM of galaxies.
}

\edit{
It is also worth appreciating that the sample sizes in FIRE and Thesan Zoom are substantially smaller than in SPICE Bursty: at $z=5$ there are 19 valid FIRE galaxies, 8 in Thesan Zoom, and 753 in SPICE Bursty.
The order of magnitude discrepancy in the number of galaxies in each sample therefore complicates efforts to make robust statements about the scatter in the measured gradients in FIRE and Thesan Zoom.
Moreover, the larger sample of SPICE Bursty also ensures a wider range of environments and merger histories than in FIRE or Thesan Zoom, which can have distinct impacts and may increase the spread of metallicity gradients.
}

\subsection{Comparison with High-Redshift (\texorpdfstring{$z>3$}{z>3}) Observations}
\label{subsec:comparison_observations}

We now make a detailed comparison to observed gas-phase metallicity gradients, with an emphasis on high-redshift $(z>3)$.
We provide a more detailed comparison with low redshift observations in the smooth feedback models in Section~4.3.1 of \cite{Garcia_2025b}.
We note that the FIRE-2 core \edit{suite} also spans $z<5$, which we did not discuss previously.
Briefly, we find the lower redshift FIRE-2 sample to be broadly consistent with the wide diversity of gradients observed at low redshift \citep[see also][for more detailed comparisons of metallicity gradients at $z<5$ in FIRE-2 galaxies]{Ma_2017,Sun_2024}.

There have been several observations of metallicity gradients at redshifts $z>3$ to date \citep[][]{Troncoso_2014,Arribas_2024,Tripodi_2024,Vallini_2024,Venturi_2024,Rodriguez_Del_Pino_2024,Acharyya_2025,Fujimoto_2025,Ivey_2025,Li_2025}, with the majority coming from {\it JWST} in the last few years.
We show these observational metallicity gradients in both Figure~\ref{fig:stellar_mass} and Figure~\ref{fig:13Gyr} (noting that Figure~\ref{fig:13Gyr} also contains observational gradients from $z<3$; which are discussed in detail in \citeauthor{Garcia_2025b} \citeyear{Garcia_2025b}).
We note that stellar mass bins do not fit so neatly into our classification in Figure~\ref{fig:stellar_mass} for \cite{Li_2025}.
We suspect, however, that the minor overlap in stellar mass does not significantly influence our comparison.
On the whole, the current consensus at high-redshift shows a preference for flatter metallicity gradients (which can be generally seen in Figure~\ref{fig:all_gradients}); however, stacked observations from \cite{Li_2025} find very strong negative gradients (which we discuss in more detail in Section~\ref{subsubsec:observational_systematics}).
This picture seems to qualitatively align with the bursty feedback models presented in this work (although it does not necessarily suggest that feedback in the observed Universe is necessarily as bursty as these simulations; see discussion in Section~\ref{subsubsec:turbulent_metals}).

We further consider the stellar mass of the observed galaxies, when available.
In our lowest mass bin ($M_\star < 10^8~{\rm M}_\odot$; top left of Figure~\ref{fig:stellar_mass}), \cite{Li_2025} have measured the gradients of stacked galaxies from $z\sim5-9$ finding strong negative gradients.
These strong gradients are most consistent with the SPICE Smooth simulations; however, as noted in Section~\ref{subsec:stellar_mass}, this mass range is not very discriminatory between the different feedback models.
As mentioned in Section~\ref{subsec:stellar_mass}, the bursty feedback models have a wider spread of gradients in this stellar mass bin.
Therefore, while the stacked gradients from \cite{Li_2025} are potentially suggestive of smooth feedback scenarios, the larger spread in bursty feedback models complicates the picture.
Observational gradients in the mass bin $10^8~{\rm M}_\odot < M_\star < 10^9~{\rm M}_\odot$ (top right panel of Figure~\ref{fig:stellar_mass}) are mostly consistent with no gradient \citep{Venturi_2024,Ivey_2025,Acharyya_2025}, with the exceptions of the strong stacked gradients at $z\sim5-9$ \citep{Li_2025} and a strongly inverted gradient at $z\sim4$ \citep{Troncoso_2014}.
As with the lowest mass bin, the observed gradients here are also suggestive.
The gradients from \cite{Venturi_2024}, \cite{Ivey_2025}, and \cite{Acharyya_2025} are most similar to those of the bursty stellar feedback models; however, the gradients are only quantitatively different to smooth feedback models, not qualitatively different.
In the mass bin $10^9~{\rm M}_\odot < M_\star < 10^{10}~{\rm M}_\odot$ (bottom left panel of Figure~\ref{fig:stellar_mass}) the few observed gradients \citep{Troncoso_2014,Venturi_2024,Li_2025} are more suggestive of bursty feedback (although the $z\sim3-4$ inverted gradients are difficult to reconcile, see discussion in Section~\ref{subsubsec:inverted_gradients} below).
The sample sizes at $z>5$ are small, but both the stacked observations and measurement of an individual galaxy are highly consistent with the flat gradients from both FIRE and Thesan Zoom, compared to the strong negative gradients in Thesan Box.

\begin{figure}
    \centering
    \includegraphics[width=\linewidth]{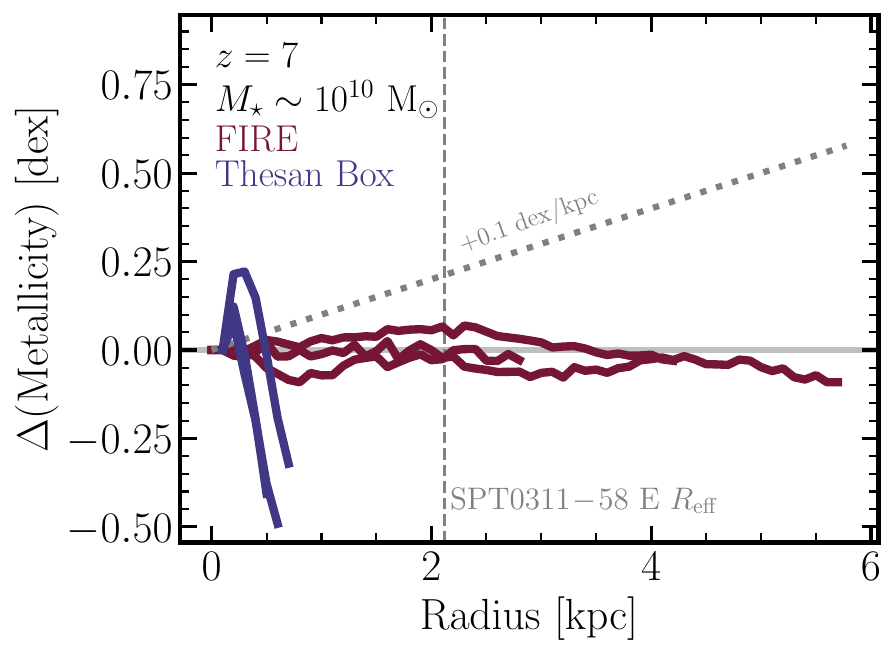}
    \caption{{\bf SPT0311-58 E Mass and Redshift Analogs in FIRE and Thesan Box.}
    Comparison of metallicity profiles from $z=7$ galaxies with stellar masses greater than $10^{10}~{\rm M}_\odot$ \protect\citep[roughly corresponding to SPT0311-58 E][]{Arribas_2024} from FIRE (dark red lines) and Thesan Box (blue lines).
    The values on the ordinate are scaled to the central metallicity value of the profile for ease of comparison.
    As a reference, we also overplot a line corresponding to a $+0.1$ dex/kpc radial gradient, which is what \protect\cite{Arribas_2024} find the gradient of SPT0311-58 E to be.
    The dashed horizontal line indicates the effective radius of SPT0311-58 E ($2.12$~kpc).
    As we discuss in Section~\ref{subsubsec:inverted_gradients}, simulation models generally struggle to reproduce strong positive gradients.
    }
    \label{fig:analogs}
\end{figure}

In the highest mass bin ($M_\star > 10^{10}~{\rm M}_\odot$), there is only one gradient quoted in the literature \citep{Arribas_2024} at $z>5$ with a few from \cite{Troncoso_2014} at $z\sim3-4$ (which we discuss in more detail in the following subsection).
Since there are only a few systems with this mass in both the observed and simulated samples at $z=7$, Figure~\ref{fig:analogs} shows the metallicity gradients of all analogs to the observed galaxy (SPT0311-58 E).
Specifically, there are six objects at $z=7$ with the same stellar mass of to SPT0311-58 E in our simulation samples: three in the FIRE high-redshift \edit{suite} (z7m12a, b, and c) and three in Thesan Box (snapshot 55 subhalo ids 0, 433, and 1612).\ignorespaces
\footnote{\ignorespaces
We note that these galaxies are selected as analogs to SPT0311-58 E purely based on their stellar mass and redshift.
It is possible that these analogs may not be direct analogs in many important ways (e.g., environment, assembly history, etc).
Moreover, there is some debate as to whether observational stellar masses can be derived robustly at these high-redshifts \citep[see, e.g.,][]{Narayanan_2024b,Cochrane_2025}.
We therefore caution against too strong an interpretation of the comparisons made between the observed sample and simulation analogs.
}
In order to avoid normalization differences, we show the 1D median metallicity profiles in Figure~\ref{fig:analogs} with respect to the central-most point of the profile.
The extent of the profile corresponds to the extent of the region we take the gradient over (which, as demonstrated in Figure~\ref{fig:gradient_comparison}, can be thought of as a proxy for the size of the galaxy).
We first note that the SPT0311-58 E analogs in Thesan Box are extremely compact, with most of their star formation taking place within the central $\sim$kpc of their centers \citep{Shen_2024b}.
The FIRE galaxies, on the other hand, are far larger, having the bulk of their star formation within $3-5$ kpc.
The Thesan Box galaxies all have slightly off-center enhancement in metallicity by $\sim0.1-0.2$ dex\ignorespaces
\footnote{\ignorespaces
The central enhancement may be caused by the kinetic mode AGN feedback in the Thesan Box model or, less physically, spatial softening lengths (which is $\sim0.2$ physical kpc at this redshift, a significant fraction of the size of the galaxies) and/or small errors in the centering of the galaxy.
}
followed by a sharp drop in the metallicity by $\sim0.4-0.6$~dex.
On the other hand, the FIRE galaxies show very little systematic variation in their median radial profiles.
The measured gradients in Thesan are thus significantly steeper ($\sim -0.5$~to~$-0.8$~dex/kpc) than in FIRE ($\sim0.0$~dex/kpc).
We note that none of the systems in FIRE or Thesan Box are particularly good analogs to SPT0311-58 E in terms of a strong positive metallicity gradient, which we discuss further in Section~\ref{subsubsec:inverted_gradients}.
Regardless, we make a specific note of the \textit{qualitatively} different nature of gradients in the smooth and bursty feedback models and suggest that further observations of massive galaxies at high-redshift would prove highly informative for the nature of stellar feedback in the earliest epoch of galaxy formation.

\edit{
Furthermore, it is worth acknowledging that a recent high redshift observations suggest that a significant fraction of systems (perhaps as high as $45\%$) show evidence of being disks \citep[e.g.,][]{Ferreira_2022,SunW_2024}.
The bursty feedback models tend to have a more chaotic, clumpy ISM due to the episodic blowouts of gas, whereas the smooth feedback models tend to form disk-like structures earlier~\citep{Shen_2024b,Shen_2025b}.
Despite the qualitative disagreement between the (lack of) structure in the bursty models, the current observations seem to favor flatter gradients.
While more careful, detailed comparisons are needed to measure the extent to which the structure in the different feedback models matches the observations, it is an intriguing tension that current models cannot, en masse, reproduce both structure at early times and flattened (or inverted) metallicity gradients.
}

\subsubsection{On Observational Systematics}
\label{subsubsec:observational_systematics}



It should be noted that there are a number of observational systematics that may influence the measure of metallicity gradients of galaxies.
Here, we briefly discuss some of the limiting factors and how they may systematically impact observed samples.

The choice of metallicity calibration, for one, can have a significant impact on the measured gradient of observed galaxies.
As a concrete example, \cite{Ivey_2025} find that the measured gradient of their galaxy ID6355 goes from $-0.01\pm0.01$ dex/kpc to $-0.11\pm 0.03$ dex/kpc with changes from strong line to direct electron temperature diagnostics, respectively.\ignorespaces
\footnote{\ignorespaces
We note that in Figures~\ref{fig:stellar_mass}~and~\ref{fig:13Gyr} we quote the strong line diagnostic gradient from \cite{Ivey_2025}.}


Another important consideration -- given the compact size of the high-redshift systems we analyze here \citep[][see also example gradients in Figure~\ref{fig:analogs}]{Shen_2024b,McClymont_2025b} -- is the impact of spatial resolution.
\cite{Yuan_2013} show that flatter gradients can be obtained just by downgrading the angular resolution of the observations.
Worse angular resolution can create a smearing effect wherein the emission lines are weighted towards regions of stronger emission, potentially causing an overestimate of the metallicity in the outskirts of the galaxy and flattening a derived gradient.
Moreover, \cite{Poetrodjojo_2019} show that the contribution of diffusion ionized gas (DIG) can also significantly flatten gradients low resolution observations.
Metallicity diagnostics are generally developed for resolutions on the order of single {\sc Hii} regions \citep{Kewley_2019}, yet current high-redshift observations are likely not resolving these small scales within the ISM.
The addition of DIG, which contributes more significantly further from the center of galaxies \citep{Kewley_2019}, can cause commonly used metallicity diagnostics to be overestimated at large radii, flattening gradients.
Careful considerations for spatial resolution are thus required to robustly measure metallicity gradients at high-redshift.
To date, several works have leveraged the power of {\it JWST} to measure metallicity gradients \citep[e.g.,][]{Arribas_2024,Venturi_2024,Acharyya_2025,Fujimoto_2025,Li_2025}, mostly finding flat gradients (with the exception of stacked gradients from \citeauthor{Li_2025} \citeyear{Li_2025}).
It is interesting to note that the stacked observations are the only measurements that give strong negative gradients at $z>3$; however, it is presently unclear if stacking gradients in this way systematically biases the result towards more negative gradients \citep[see][for a more complete discussion of their stacking procedure]{Li_2025}.
Regardless, future high-resolution efforts that make considerations for the spatial resolution and contributions from DIG to obtain metallicity gradients at high-redshift could potentially help discriminate between the different feedback models.

Another consideration is that, in the smooth feedback models, the star-forming regions of galaxies at high-redshift are highly compact (see Figure~\ref{fig:analogs}) with strong negative gradients that flatten significantly outside this region (see Figure~\ref{fig:gradient_comparison} for an example in Thesan Box; see also \citeauthor{Garcia_2023} \citeyear{Garcia_2023}, \citeauthor{Tapia_2025} \citeyear{Tapia_2025}).
This gradient flattening in the smooth feedback models is set by the competition between gas mixing and enrichment at large radii \citep{Garcia_2023}.
Observations of massive galaxies with poor spatial resolution could potentially wash out the small, high metallicity interiors of these galaxies and find a flatter gradient than we report here.
The extent to which, e.g., the SPT0311-58 E analogs (discussed in Section~\ref{subsubsec:inverted_gradients}) would be discriminatory for feedback physics in the early Universe is likely dependent on metallicity diagnostic and spatial resolution of the observations.

Finally, we note that we rotate our galaxies to the face-on orientation (see Section~\ref{subsec:gradient_definitions}).
While not so physically meaningful for turbulent, assembling galaxies in bursty feedback models, this could play a role in the measured gradient of galaxies in a smooth feedback scenario.
Many low-redshift studies estimate the inclination of the disk and deproject it to the face-on orientation \citep[][etc]{Grasha_2022,Vale_2025}.
This practice is not as common at higher redshift, where the exact orientation of a disk (if, indeed, there is one) would not necessarily be known \citep[e.g.,][who make no mention of deprojection efforts]{Vallini_2024,Venturi_2024,Acharyya_2025,Fujimoto_2025,Li_2025}.
In a bursty feedback scenario, since the galaxies have highly turbulent, clumpy structure, it \edit{is} unlikely that the lack of deprojection plays a significantly role; however, in smooth feedback models it is possible that viewing a disk galaxy at some orientation could significantly impact the derived gradient.

We therefore caution that, while tantalizing that the majority of observations align more closely with the bursty feedback models, the extent to which conclusions can be robustly drawn given the current observations is presently unclear.

\subsubsection{On Simulation Modeling Limitations}
\label{subsubsec:turbulent_metals}

If the observations of flattened gradients are robust to the above systematics, there is evidence that the smooth feedback models may not sufficiently mix the metal content within their ISM.
While rapid expulsions of gas driven by bursty stellar feedback is one method by which this mixing can be driven, it is not necessarily the only method.
For example, diffusion of metals from unresolved turbulence within the ISM could play a significant role in redistributing metals and flattening gradients.
Using the FIRE-2 model for this unresolved turbulence \citep[described in][]{Su_2017,Escala_2018,Hopkins_2018}, \cite{Bellardini_2021} show that changes to the diffusion coefficients can systematically impact the radial distribution of metals.
Specifically, those authors find that increased metal diffusion flattens metallicity gradients.
It is therefore possible that the smooth feedback models require increased metal mixing from these unresolved eddies.

Increased turbulent diffusion mixes metals, but likely would not significantly change the overall total metal content.
The existence---or lack thereof---of a correlation between the scatter about the mass-metallicity relation (MZR) and star formation rates (often referred to as the Fundamental Metallicity Relation; FMR) would therefore likely also be unimpeded in smooth feedback models with increased subgrid metal diffusion.
While there is evidence that the FMR evolves in the smooth feedback models (i.e., the FMRs are weak and/or dynamic; \citeauthor{Garcia_2024b} \citeyear{Garcia_2024b}, \citeyear{Garcia_2025a}), there still exists an anti-correlation between metallicity and SFR up to at least $z=8$.
Bursty feedback, on the other hand, should disrupt the interplay of gas accretion and metal return that gives rise to this anti-correlation \citep{Garcia_2024a,Garcia_2024b,Garcia_2025a,Bassini_2024,McClymont_2025}.
The existence of FMR-like anti-correlations between metallicity and SFR in the high-redshift Universe---regardless of strong/weak or static/dynamic designation---could provide further constraining power as to whether flattened gradients are suggestive of bursty feedback or increased small-scale turbulent eddies within the ISM.


\subsubsection{On the Dearth of Strong Positive Gradients in Simulations}
\label{subsubsec:inverted_gradients}

There are three main ways proposed to create a strong positive (sometimes called inverted) metallicity gradient: {\it (i)} galaxy-galaxy mergers/interactions \citep[e.g.,][]{Rupke_2010a,Torrey_2012,Grossi_2020}, {\it (ii)} pristine gas accretion onto the central regions of a galaxy \citep[e.g.,][]{Cresci_2010}, and/or {\it (iii)} strong outflows \citep[e.g.,][]{Rodriguez_Del_Pino_2024}.
While it is common to preferentially omit merging systems in low redshift zoom-ins, neither Thesan Zoom or the high-redshift FIRE-2 runs used here have such a restriction.
Furthermore, a bias against mergers is likely not true for the volumes (SPICE Smooth/Bursty, Thesan Box, EAGLE, Illustris, IllustrisTNG, and SIMBA) which tend to select regions where the average overdensity is approximately 0.
It is also unlikely that the full volume samples preferentially select against galaxies that have gas accretion.
We also know that the bursty feedback models do drive strong outflows of metals \citep{Muratov_2015,Muratov_2017}.
Yet, despite our models nominally including the impact of all three proposed mechanisms required to invert gradients, there are very few galaxies in our entire sample (only $\sim1.4\%$) that exhibit gradients greater than $+0.1$~dex/kpc, and are particularly rare at larger stellar masses and in smooth feedback models.

It is interesting to note that observations from \cite{Troncoso_2014} present a whole population of strong positive metallicity gradients for ten galaxies at $M_\star > 10^{9}~{\rm M}_\odot$ at $z\sim3-4$.
The gradients of these intermediate-to-high mass galaxies are difficult to explain with stellar feedback alone.
Even our bursty feedback models do not have a significant population of high mass galaxies with inverted gradients.
While there is some evidence that merging/interacting galaxies can produce strong positive metallicity gradients, \cite{Troncoso_2014} suggest that the population of inverted gradients they observe at $z=3-4$ are likely not mergers.
Instead, \cite{Troncoso_2014} suggest that the strong positive gradient could be caused by strong pristine gas accretion into the central regions of the galaxy.

Put more quantitatively, we overplot a dashed line of $+0.1$~dex/kpc on Figure~\ref{fig:analogs}, corresponding to the measured gradient of SPT0311-58 E \citep{Arribas_2024}.
While the Thesan Box and FIRE gradients have qualitatively different behavior (see discussion above), both models have significantly different gradients than the observed $+0.1$~dex/kpc gradient of SPT0311-58 E.
Short of systematic effects with spatial resolution or metallicity diagnostics (see discussion in Section~\ref{subsubsec:observational_systematics}), it is unclear how to explain the strong positive gradient observed in SPT0311-58 E.

\edit{
If the measurements of positive gradients are robust, one possibility is that the bursty feedback models {\it over}-mix their metal content.
In the previous section and \cite{Garcia_2025b}, we argue that many smooth feedback models may undermix their metals due to the lack of a subgrid turbulent diffusion model.
It is possible that the bursty feedback models overcorrect the metal mixing, washing out any gradient--positive or negative--either through subgrid turbulence models or overly strong stellar feedback-driven outflows (or a combination thereof; see also discussion in Section~\ref{subsubsec:bursty_model_comparisons}).
There may be a ``sweet spot'' between the burstiness of feedback and metal diffusion coefficient that allows gradients to persist, but not be (virtually) uniformly negative.
The existence and subsequent persistence of metallicity gradients therefore also can place strong constraints on the efficiency of metal mixing within the ISM required for future simulation efforts \cite[see also][]{Hirai_2017,Escala_2018}.
}

In summary, the majority of gradients are either relatively flat or strongly negative, regardless of feedback implementation.
The population of strong positive gradients seen in observations across redshift can therefore not be easily explained in the scatter of current simulations models.

\section{Conclusions}
\label{sec:conclusions}

In this work, we analyze the gas-phase radial metallicity gradients in the modern high-redshift ($3<z\lesssim11$) cosmological simulations FIRE, SPICE, and Thesan.
This suite of models is particularly advantageous because it spans a range of star formation modes---both strongly time-variable (i.e., bursty) and relatively smooth (see Figure~\ref{fig:SFH_plot})---and therefore captures the corresponding diversity in stellar feedback.
We construct one dimensional gas-phase metallicity profiles for each galaxy and fit it with a linear regression (see Figure~\ref{fig:gradient_comparison} for examples from each simulation model).

Our key conclusions are as follows:
\begin{itemize}[leftmargin=0pt]
    \item We find that metallicity gradients are systematically flatter (factors of $\sim2-10$) in bursty  (FIRE-2, SPICE Bursty, and Thesan Zoom) than in smooth  (SPICE Smooth and Thesan Box) feedback models (Figure~\ref{fig:all_gradients}).
    We note that this result is---remarkably---mostly independent of the detailed implementations of either bursty or smooth feedback.

    \item We find that the discrepancy between smooth and bursty feedback models is most clear in massive galaxies ($M_\star > 10^{9}~{\rm M}_\odot$; Figure~\ref{fig:stellar_mass}).
    Low mass galaxies do show some level of variation between the different feedback implementations, but the difference is more subtle and would likely be difficult to discriminate with the current data.

    \item We contextualize our results with the first paper in this series \citep{Garcia_2025b}, which shows smooth feedback models at $0\leq z\leq8$, and recent {\it JWST}/ALMA observations (Figure~\ref{fig:13Gyr}).
    We find good agreement between the metallicity gradients in smooth feedback models of this work (Thesan Box and SPICE Smooth) and EAGLE, Illustris, TNG, and SIMBA, suggesting that the strong negative gradients are a highly generic feature of smooth feedback models.
    Moreover, the agreement between FIRE, SPICE Bursty, and Thesan Zoom suggests that the flat gradients may be a generic feature of bursty feedback models.

    \item Finally, we find that the recent high-redshift gradient observations \citep{Arribas_2024,Vallini_2024,Venturi_2024,Li_2025} tend to align more closely with the gradients from bursty feedback models.
    This is suggestive that smooth feedback models may under-mix their metal content, either through increased feedback or a lack of a metal diffusion in the models (see discussion in Section~\ref{subsubsec:turbulent_metals}); however, given current observational limitations, it is unclear how robust this finding is (see discussion in Section~\ref{subsubsec:observational_systematics}).
    Moreover, we find that both smooth and bursty models struggle to reproduce strong positive gradients at all (with only $\sim1.5\%$ of the entire sample having gradients $>0.1~{\rm dex/kpc}$; again, with the above caveat about observational limitations).
\end{itemize}
Already {\it JWST} has transformed our understanding of the role of stellar feedback in the early Universe.
The results presented in this work and \cite{Garcia_2025b} further demonstrate the utility for continued observations of metallicity gradients at high-redshift to be constraining for the nature of feedback.
Smooth and bursty feedback models make {\it qualitatively} different predictions for the strength, and evolution, of gradients.
A large sample of high resolution observations of galaxies in the first few $100$ Myrs of the history of the Universe should provide strong constraints for our understanding of stellar feedback.

\begin{acknowledgments}

AMG acknowledges support from a Virginia Space Grant Consortium Graduate STEM Research Fellowship.
AMG and PT acknowledge support from NSF-AST 2346977 and the NSF-Simons AI Institute for Cosmic Origins which is supported by the National Science Foundation under Cooperative Agreement 2421782 and the Simons Foundation award MPS-AI-00010515.
XS acknowledges support from the NASA theory grant JWST-AR-04814.
WM thanks the Science and Technology Facilities Council (STFC) Center for Doctoral Training (CDT) in Data Intensive Science at the University of Cambridge (STFC grant number 2742968) for a PhD studentship. WM acknowledges support by the Royal Society Research Grant G125142.
LK acknowledges the support of a Royal Society University Research Fellowship (grant number URF$\backslash$R1$\backslash$251793).
RK acknowledges support of the Natural Sciences and Engineering Research Council of Canada (NSERC) through a Discovery Grant and a Discovery Launch Supplement (funding reference numbers RGPIN-2024-06222 and DGECR-2024-00144) and York University's Global Research Excellence Initiative.

The authors acknowledge Research Computing at The University of Virginia, University of Florida Information Technology, Massachusetts Institute of Technology, and Max Planck Computing and Data Facility for providing computational resources and technical support that have contributed to the results reported within this publication.
\end{acknowledgments}

\bibliography{references}{}

\begin{thebibliography}{}
\expandafter\ifx\csname natexlab\endcsname\relax\def\natexlab#1{#1}\fi
\providecommand{\url}[1]{\href{#1}{#1}}
\providecommand{\dodoi}[1]{doi:~\href{http://doi.org/#1}{\nolinkurl{#1}}}
\providecommand{\doeprint}[1]{\href{http://ascl.net/#1}{\nolinkurl{http://ascl.net/#1}}}
\providecommand{\doarXiv}[1]{\href{https://arxiv.org/abs/#1}{\nolinkurl{https://arxiv.org/abs/#1}}}

\bibitem[{{Acharyya} {et~al.}(2020){Acharyya}, {Krumholz}, {Federrath}, {Kewley}, {Goldbaum}, \& {Sharp}}]{Acharyya_2020}
{Acharyya}, A., {Krumholz}, M.~R., {Federrath}, C., {et~al.} 2020, \mnras, 495, 3819, \dodoi{10.1093/mnras/staa1100}

\bibitem[{{Acharyya} {et~al.}(2025){Acharyya}, {Watson}, {Vulcani}, {Treu}, {Nedkova}, {Bunker}, {Mehta}, {Atek}, {Battisti}, {Hasan}, {Hayes}, {Huberty}, {Jones}, {Leethochawalit}, {Lin}, {Malkan}, {Metha}, {Nanayakkara}, {Rafelski}, {Sattari}, {Scarlata}, {Wang}, {Casey}, {Grazian}, {Koekemoer}, {Radovich}, \& {Rodighiero}}]{Acharyya_2025}
{Acharyya}, A., {Watson}, P.~J., {Vulcani}, B., {et~al.} 2025, arXiv e-prints, arXiv:2508.05335, \dodoi{10.48550/arXiv.2508.05335}

\bibitem[{{Agertz} {et~al.}(2011){Agertz}, {Teyssier}, \& {Moore}}]{Agertz_2011}
{Agertz}, O., {Teyssier}, R., \& {Moore}, B. 2011, \mnras, 410, 1391, \dodoi{10.1111/j.1365-2966.2010.17530.x}

\bibitem[{{Angl{\'e}s-Alc{\'a}zar} {et~al.}(2017){Angl{\'e}s-Alc{\'a}zar}, {Faucher-Gigu{\`e}re}, {Kere{\v{s}}}, {Hopkins}, {Quataert}, \& {Murray}}]{Angles_Alcazar_2017b}
{Angl{\'e}s-Alc{\'a}zar}, D., {Faucher-Gigu{\`e}re}, C.-A., {Kere{\v{s}}}, D., {et~al.} 2017, \mnras, 470, 4698, \dodoi{10.1093/mnras/stx1517}

\bibitem[{{Arribas} {et~al.}(2024){Arribas}, {Perna}, {Rodr{\'\i}guez Del Pino}, {Lamperti}, {D'Eugenio}, {P{\'e}rez-Gonz{\'a}lez}, {Jones}, {Crespo G{\'o}mez}, {Curti}, {Lim}, {{\'A}lvarez-M{\'a}rquez}, {Bunker}, {Carniani}, {Charlot}, {Jakobsen}, {Maiolino}, {{\"U}bler}, {Willott}, {B{\"o}ker}, {Chevallard}, {Circosta}, {Cresci}, {Kumari}, {Parlanti}, {Scholtz}, {Venturi}, \& {Witstok}}]{Arribas_2024}
{Arribas}, S., {Perna}, M., {Rodr{\'\i}guez Del Pino}, B., {et~al.} 2024, \aap, 688, A146, \dodoi{10.1051/0004-6361/202348824}

\bibitem[{{Bassini} {et~al.}(2024){Bassini}, {Feldmann}, {Gensior}, {Faucher-Gigu{\`e}re}, {Cenci}, {Moreno}, {Bernardini}, \& {Liang}}]{Bassini_2024}
{Bassini}, L., {Feldmann}, R., {Gensior}, J., {et~al.} 2024, \mnras, 532, L14, \dodoi{10.1093/mnrasl/slae036}

\bibitem[{{Belfiore} {et~al.}(2017){Belfiore}, {Maiolino}, {Tremonti}, {S{\'a}nchez}, {Bundy}, {Bershady}, {Westfall}, {Lin}, {Drory}, {Boquien}, {Thomas}, \& {Brinkmann}}]{Belfiore_2017}
{Belfiore}, F., {Maiolino}, R., {Tremonti}, C., {et~al.} 2017, \mnras, 469, 151, \dodoi{10.1093/mnras/stx789}

\bibitem[{{Bellardini} {et~al.}(2021){Bellardini}, {Wetzel}, {Loebman}, {Faucher-Gigu{\`e}re}, {Ma}, \& {Feldmann}}]{Bellardini_2021}
{Bellardini}, M.~A., {Wetzel}, A., {Loebman}, S.~R., {et~al.} 2021, \mnras, 505, 4586, \dodoi{10.1093/mnras/stab1606}

\bibitem[{{Bhagwat} {et~al.}(2024){Bhagwat}, {Costa}, {Ciardi}, {Pakmor}, \& {Garaldi}}]{Bhagwat_2024}
{Bhagwat}, A., {Costa}, T., {Ciardi}, B., {Pakmor}, R., \& {Garaldi}, E. 2024, \mnras, 531, 3406, \dodoi{10.1093/mnras/stae1125}

\bibitem[{{Brook} {et~al.}(2012){Brook}, {Stinson}, {Gibson}, {Wadsley}, \& {Quinn}}]{Brook_2012}
{Brook}, C.~B., {Stinson}, G., {Gibson}, B.~K., {Wadsley}, J., \& {Quinn}, T. 2012, \mnras, 424, 1275, \dodoi{10.1111/j.1365-2966.2012.21306.x}

\bibitem[{{Burger} {et~al.}(2025){Burger}, {Springel}, {Ostriker}, {Kim}, {Jeffreson}, {Smith}, {Pakmor}, {Hassan}, {Fielding}, {Hernquist}, {Bryan}, {Somerville}, {Bennett}, \& {Weinberger}}]{Burger_2025}
{Burger}, J.~D., {Springel}, V., {Ostriker}, E.~C., {et~al.} 2025, \mnras, \dodoi{10.1093/mnras/staf1720}

\bibitem[{{Carton} {et~al.}(2018){Carton}, {Brinchmann}, {Contini}, {Epinat}, {Finley}, {Richard}, {Patr{\'\i}cio}, {Schaye}, {Nanayakkara}, {Weilbacher}, \& {Wisotzki}}]{Carton_2018}
{Carton}, D., {Brinchmann}, J., {Contini}, T., {et~al.} 2018, \mnras, 478, 4293, \dodoi{10.1093/mnras/sty1343}

\bibitem[{{Ceverino} {et~al.}(2016){Ceverino}, {S{\'a}nchez Almeida}, {Mu{\~n}oz Tu{\~n}{\'o}n}, {Dekel}, {Elmegreen}, {Elmegreen}, \& {Primack}}]{Ceverino_2016}
{Ceverino}, D., {S{\'a}nchez Almeida}, J., {Mu{\~n}oz Tu{\~n}{\'o}n}, C., {et~al.} 2016, \mnras, 457, 2605, \dodoi{10.1093/mnras/stw064}

\bibitem[{Chabrier(2003)}]{Chabrier_2003}
Chabrier, G. 2003, PASP, 115, 763, \dodoi{10.1086/376392}

\bibitem[{{Chan} {et~al.}(2018){Chan}, {Kere{\v{s}}}, {Wetzel}, {Hopkins}, {Faucher-Gigu{\`e}re}, {El-Badry}, {Garrison-Kimmel}, \& {Boylan-Kolchin}}]{Chan_2018}
{Chan}, T.~K., {Kere{\v{s}}}, D., {Wetzel}, A., {et~al.} 2018, \mnras, 478, 906, \dodoi{10.1093/mnras/sty1153}

\bibitem[{{Cochrane} {et~al.}(2025){Cochrane}, {Katz}, {Begley}, {Hayward}, \& {Best}}]{Cochrane_2025}
{Cochrane}, R.~K., {Katz}, H., {Begley}, R., {Hayward}, C.~C., \& {Best}, P.~N. 2025, \apjl, 978, L42, \dodoi{10.3847/2041-8213/ad9a4d}

\bibitem[{{Crain} \& {van de Voort}(2023)}]{Crain_2023}
{Crain}, R.~A., \& {van de Voort}, F. 2023, \araa, 61, 473, \dodoi{10.1146/annurev-astro-041923-043618}

\bibitem[{{Cresci} {et~al.}(2010){Cresci}, {Mannucci}, {Maiolino}, {Marconi}, {Gnerucci}, \& {Magrini}}]{Cresci_2010}
{Cresci}, G., {Mannucci}, F., {Maiolino}, R., {et~al.} 2010, \nat, 467, 811, \dodoi{10.1038/nature09451}

\bibitem[{{Curti} {et~al.}(2020){Curti}, {Maiolino}, {Cirasuolo}, {Mannucci}, {Williams}, {Auger}, {Mercurio}, {Hayden-Pawson}, {Cresci}, {Marconi}, {Belfiore}, {Cappellari}, {Cicone}, {Cullen}, {Meneghetti}, {Ota}, {Peng}, {Pettini}, {Swinbank}, \& {Troncoso}}]{Curti_2020}
{Curti}, M., {Maiolino}, R., {Cirasuolo}, M., {et~al.} 2020, \mnras, 492, 821, \dodoi{10.1093/mnras/stz3379}

\bibitem[{{Dale} {et~al.}(2005){Dale}, {Bonnell}, {Clarke}, \& {Bate}}]{Dale_2005}
{Dale}, J.~E., {Bonnell}, I.~A., {Clarke}, C.~J., \& {Bate}, M.~R. 2005, \mnras, 358, 291, \dodoi{10.1111/j.1365-2966.2005.08806.x}

\bibitem[{{Danhaive} {et~al.}(2025){Danhaive}, {Tacchella}, {\textbackslash''Ubler}, {de Graaff}, {Egami}, {Johnson}, {Sun}, {Arribas}, {Bunker}, {Carniani}, {Jones}, {Maiolino}, {McClymont}, {Parlanti}, {Simmonds}, {Villanueva}, {Baker}, {Jaffe}, {Eisenstein}, {Hainline}, {Helton}, {Ji}, {Lin}, {Pusk\textbackslash'as}, {Rieke}, {Rinaldi}, {Robertson}, {Scholz}, {Williams}, \& {Willmer}}]{Danhaive_2025}
{Danhaive}, A.~L., {Tacchella}, S., {\textbackslash''Ubler}, H., {et~al.} 2025, arXiv e-prints, arXiv:2503.21863, \dodoi{10.48550/arXiv.2503.21863}

\bibitem[{{Dav{\'e}} {et~al.}(2019){Dav{\'e}}, {Angl{\'e}s-Alc{\'a}zar}, {Narayanan}, {Li}, {Rafieferantsoa}, \& {Appleby}}]{Dave_2019}
{Dav{\'e}}, R., {Angl{\'e}s-Alc{\'a}zar}, D., {Narayanan}, D., {et~al.} 2019, \mnras, 486, 2827, \dodoi{10.1093/mnras/stz937}

\bibitem[{{Doherty} {et~al.}(2014){Doherty}, {Gil-Pons}, {Lau}, {Lattanzio}, \& {Siess}}]{Doherty_2014}
{Doherty}, C.~L., {Gil-Pons}, P., {Lau}, H. H.~B., {Lattanzio}, J.~C., \& {Siess}, L. 2014, \mnras, 437, 195, \dodoi{10.1093/mnras/stt1877}

\bibitem[{{El-Badry} {et~al.}(2016){El-Badry}, {Wetzel}, {Geha}, {Hopkins}, {Kere{\v{s}}}, {Chan}, \& {Faucher-Gigu{\`e}re}}]{El_Badry_2016}
{El-Badry}, K., {Wetzel}, A., {Geha}, M., {et~al.} 2016, \apj, 820, 131, \dodoi{10.3847/0004-637X/820/2/131}

\bibitem[{{El-Badry} {et~al.}(2018){El-Badry}, {Quataert}, {Wetzel}, {Hopkins}, {Weisz}, {Chan}, {Fitts}, {Boylan-Kolchin}, {Kere{\v{s}}}, {Faucher-Gigu{\`e}re}, \& {Garrison-Kimmel}}]{El_Badry_2018}
{El-Badry}, K., {Quataert}, E., {Wetzel}, A., {et~al.} 2018, \mnras, 473, 1930, \dodoi{10.1093/mnras/stx2482}

\bibitem[{{Escala} {et~al.}(2018){Escala}, {Wetzel}, {Kirby}, {Hopkins}, {Ma}, {Wheeler}, {Kere{\v{s}}}, {Faucher-Gigu{\`e}re}, \& {Quataert}}]{Escala_2018}
{Escala}, I., {Wetzel}, A., {Kirby}, E.~N., {et~al.} 2018, \mnras, 474, 2194, \dodoi{10.1093/mnras/stx2858}

\bibitem[{{Evans} {et~al.}(2009){Evans}, {Dunham}, {J{\o}rgensen}, {Enoch}, {Mer{\'\i}n}, {van Dishoeck}, {Alcal{\'a}}, {Myers}, {Stapelfeldt}, {Huard}, {Allen}, {Harvey}, {van Kempen}, {Blake}, {Koerner}, {Mundy}, {Padgett}, \& {Sargent}}]{Evans_2009}
{Evans}, Neal~J., I., {Dunham}, M.~M., {J{\o}rgensen}, J.~K., {et~al.} 2009, \apjs, 181, 321, \dodoi{10.1088/0067-0049/181/2/321}

\bibitem[{{Federrath} \& {Klessen}(2012)}]{Federrath_2012}
{Federrath}, C., \& {Klessen}, R.~S. 2012, \apj, 761, 156, \dodoi{10.1088/0004-637X/761/2/156}

\bibitem[{{Feldmann} {et~al.}(2023){Feldmann}, {Quataert}, {Faucher-Gigu{\`e}re}, {Hopkins}, {{\c{C}}atmabacak}, {Kere{\v{s}}}, {Bassini}, {Bernardini}, {Bullock}, {Cenci}, {Gensior}, {Liang}, {Moreno}, \& {Wetzel}}]{Feldmann_2023}
{Feldmann}, R., {Quataert}, E., {Faucher-Gigu{\`e}re}, C.-A., {et~al.} 2023, \mnras, 522, 3831, \dodoi{10.1093/mnras/stad1205}

\bibitem[{{Ferreira} {et~al.}(2022){Ferreira}, {Adams}, {Conselice}, {Sazonova}, {Austin}, {Caruana}, {Ferrari}, {Verma}, {Trussler}, {Broadhurst}, {Diego}, {Frye}, {Pascale}, {Wilkins}, {Windhorst}, \& {Zitrin}}]{Ferreira_2022}
{Ferreira}, L., {Adams}, N., {Conselice}, C.~J., {et~al.} 2022, \apjl, 938, L2, \dodoi{10.3847/2041-8213/ac947c}

\bibitem[{{Ferrero} {et~al.}(2017){Ferrero}, {Navarro}, {Abadi}, {Sales}, {Bower}, {Crain}, {Frenk}, {Schaller}, {Schaye}, \& {Theuns}}]{Ferrero_2017}
{Ferrero}, I., {Navarro}, J.~F., {Abadi}, M.~G., {et~al.} 2017, \mnras, 464, 4736, \dodoi{10.1093/mnras/stw2691}

\bibitem[{{Fishlock} {et~al.}(2014){Fishlock}, {Karakas}, {Lugaro}, \& {Yong}}]{Fishlock_2014}
{Fishlock}, C.~K., {Karakas}, A.~I., {Lugaro}, M., \& {Yong}, D. 2014, \apj, 797, 44, \dodoi{10.1088/0004-637X/797/1/44}

\bibitem[{{Forouhar Moreno} {et~al.}(2025){Forouhar Moreno}, {Helly}, {McGibbon}, {Schaye}, {Schaller}, {Han}, {Kugel}, \& {Bah{\'e}}}]{Forouhar_2025}
{Forouhar Moreno}, V.~J., {Helly}, J., {McGibbon}, R., {et~al.} 2025, \mnras, \dodoi{10.1093/mnras/staf1478}

\bibitem[{{F{\"o}rster Schreiber} {et~al.}(2018){F{\"o}rster Schreiber}, {Renzini}, {Mancini}, {Genzel}, {Bouch{\'e}}, {Cresci}, {Hicks}, {Lilly}, {Peng}, {Burkert}, {Carollo}, {Cimatti}, {Daddi}, {Davies}, {Genel}, {Kurk}, {Lang}, {Lutz}, {Mainieri}, {McCracken}, {Mignoli}, {Naab}, {Oesch}, {Pozzetti}, {Scodeggio}, {Shapiro Griffin}, {Shapley}, {Sternberg}, {Tacchella}, {Tacconi}, {Wuyts}, \& {Zamorani}}]{Forster_Schreiber_2018}
{F{\"o}rster Schreiber}, N.~M., {Renzini}, A., {Mancini}, C., {et~al.} 2018, \apjs, 238, 21, \dodoi{10.3847/1538-4365/aadd49}

\bibitem[{{Franchetto} {et~al.}(2021){Franchetto}, {Mingozzi}, {Poggianti}, {Vulcani}, {Bacchini}, {Gullieuszik}, {Moretti}, {Tomicic}, \& {Fritz}}]{Franchetto_2021}
{Franchetto}, A., {Mingozzi}, M., {Poggianti}, B.~M., {et~al.} 2021, arXiv e-prints, arXiv:2109.02656.
\newblock \doarXiv{2109.02656}

\bibitem[{{Friedli} \& {Benz}(1995)}]{Friedli_1995}
{Friedli}, D., \& {Benz}, W. 1995, \aap, 301, 649

\bibitem[{{Friedli} {et~al.}(1994){Friedli}, {Benz}, \& {Kennicutt}}]{Friedli_1994}
{Friedli}, D., {Benz}, W., \& {Kennicutt}, R. 1994, \apjl, 430, L105, \dodoi{10.1086/187449}

\bibitem[{{Fujimoto} {et~al.}(2025){Fujimoto}, {Faisst}, {Tsujita}, {Kohandel}, {Lee}, {{\"U}bler}, {Loiacono}, {Nezhad}, {Pallottini}, {Aravena}, {Assef}, {Battisti}, {B{\'e}thermin}, {Boquien}, {da Cunha}, {Ferrara}, {Franco}, {Ginolfi}, {Hadi}, {Haghjoo}, {Herrera-Camus}, {Inami}, {Koekemoer}, {Lemaux}, {Li}, {Liu}, {Molina}, {Nanni}, {Pozzi}, {Relano}, {Romano}, {Sanders}, {F{\"o}rster Schreiber}, {Silverman}, {Spilker}, {Telikova}, {Villanueva}, {Vallini}, {Wang}, \& {Zamorani}}]{Fujimoto_2025}
{Fujimoto}, S., {Faisst}, A.~L., {Tsujita}, A., {et~al.} 2025, arXiv e-prints, arXiv:2510.16116.
\newblock \doarXiv{2510.16116}

\bibitem[{{Garaldi} {et~al.}(2022){Garaldi}, {Kannan}, {Smith}, {Springel}, {Pakmor}, {Vogelsberger}, \& {Hernquist}}]{Garaldi_2022}
{Garaldi}, E., {Kannan}, R., {Smith}, A., {et~al.} 2022, \mnras, 512, 4909, \dodoi{10.1093/mnras/stac257}

\bibitem[{{Garaldi} {et~al.}(2024){Garaldi}, {Kannan}, {Smith}, {Borrow}, {Vogelsberger}, {Pakmor}, {Springel}, {Hernquist}, {Gal{\'a}rraga-Espinosa}, {Yeh}, {Shen}, {Xu}, {Neyer}, {Spina}, {Almualla}, \& {Zhao}}]{Garaldi_2024}
---. 2024, \mnras, 530, 3765, \dodoi{10.1093/mnras/stae839}

\bibitem[{{Garcia} {et~al.}(2023){Garcia}, {Torrey}, {Hemler}, {Hernquist}, {Kewley}, {Nelson}, {Grasha}, {Zovaro}, \& {Chen}}]{Garcia_2023}
{Garcia}, A.~M., {Torrey}, P., {Hemler}, Z.~S., {et~al.} 2023, \mnras, 519, 4716, \dodoi{10.1093/mnras/stac3749}

\bibitem[{{Garcia} {et~al.}(2024{\natexlab{a}}){Garcia}, {Torrey}, {Grasha}, {Hernquist}, {Ellison}, {Zovaro}, {Hemler}, {Nelson}, \& {Kewley}}]{Garcia_2024a}
{Garcia}, A.~M., {Torrey}, P., {Grasha}, K., {et~al.} 2024{\natexlab{a}}, \mnras, 529, 3342, \dodoi{10.1093/mnras/stae737}

\bibitem[{{Garcia} {et~al.}(2024{\natexlab{b}}){Garcia}, {Torrey}, {Ellison}, {Grasha}, {Hernquist}, {Zovaro}, {Chen}, {Hemler}, {Kewley}, {Nelson}, \& {Wright}}]{Garcia_2024b}
{Garcia}, A.~M., {Torrey}, P., {Ellison}, S., {et~al.} 2024{\natexlab{b}}, \mnras, 531, 1398, \dodoi{10.1093/mnras/stae1252}

\bibitem[{{Garcia} {et~al.}(2025{\natexlab{a}}){Garcia}, {Torrey}, {Ellison}, {Grasha}, {Chen}, {Hemler}, {Zimmerman}, {Wright}, {Zovaro}, {Nelson}, {Sanders}, {Kewley}, \& {Hernquist}}]{Garcia_2025a}
{Garcia}, A.~M., {Torrey}, P., {Ellison}, S.~L., {et~al.} 2025{\natexlab{a}}, \mnras, 536, 119, \dodoi{10.1093/mnras/stae2587}

\bibitem[{{Garcia} {et~al.}(2025{\natexlab{b}}){Garcia}, {Torrey}, {Bhagwat}, {Wright}, {Chen}, {Grasha}, {Ridolfo}, {Hemler}, {Sarkar}, {Chakraborty}, {Nelson}, {Sanders}, {Costa}, {Vogelsberger}, {Kewley}, {Ellison}, \& {Hernquist}}]{Garcia_2025b}
{Garcia}, A.~M., {Torrey}, P., {Bhagwat}, A., {et~al.} 2025{\natexlab{b}}, \apj, 989, 147, \dodoi{10.3847/1538-4357/adea51}

\bibitem[{{Garrison-Kimmel} {et~al.}(2017){Garrison-Kimmel}, {Wetzel}, {Bullock}, {Hopkins}, {Boylan-Kolchin}, {Faucher-Gigu{\`e}re}, {Kere{\v{s}}}, {Quataert}, {Sanderson}, {Graus}, \& {Kelley}}]{Garrison_Kimmel_2017}
{Garrison-Kimmel}, S., {Wetzel}, A., {Bullock}, J.~S., {et~al.} 2017, \mnras, 471, 1709, \dodoi{10.1093/mnras/stx1710}

\bibitem[{{Garrison-Kimmel} {et~al.}(2019){Garrison-Kimmel}, {Hopkins}, {Wetzel}, {Bullock}, {Boylan-Kolchin}, {Kere{\v{s}}}, {Faucher-Gigu{\`e}re}, {El-Badry}, {Lamberts}, {Quataert}, \& {Sanderson}}]{Garrison_Kimmel_2019}
{Garrison-Kimmel}, S., {Hopkins}, P.~F., {Wetzel}, A., {et~al.} 2019, \mnras, 487, 1380, \dodoi{10.1093/mnras/stz1317}

\bibitem[{{Gelli} {et~al.}(2024){Gelli}, {Mason}, \& {Hayward}}]{Gelli_2024}
{Gelli}, V., {Mason}, C., \& {Hayward}, C.~C. 2024, \apj, 975, 192, \dodoi{10.3847/1538-4357/ad7b36}

\bibitem[{{Genel} {et~al.}(2014){Genel}, {Vogelsberger}, {Springel}, {Sijacki}, {Nelson}, {Snyder}, {Rodriguez-Gomez}, {Torrey}, \& {Hernquist}}]{Genel_2014}
{Genel}, S., {Vogelsberger}, M., {Springel}, V., {et~al.} 2014, \mnras, 445, 175, \dodoi{10.1093/mnras/stu1654}

\bibitem[{{Gibson} {et~al.}(2013){Gibson}, {Pilkington}, {Brook}, {Stinson}, \& {Bailin}}]{Gibson_2013}
{Gibson}, B.~K., {Pilkington}, K., {Brook}, C.~B., {Stinson}, G.~S., \& {Bailin}, J. 2013, \aap, 554, A47, \dodoi{10.1051/0004-6361/201321239}

\bibitem[{{Graf} {et~al.}(2024){Graf}, {Wetzel}, {Bailin}, \& {Orr}}]{Graf_2024}
{Graf}, R.~L., {Wetzel}, A., {Bailin}, J., \& {Orr}, M.~E. 2024, arXiv e-prints, arXiv:2410.21377, \dodoi{10.48550/arXiv.2410.21377}

\bibitem[{{Grasha} {et~al.}(2022){Grasha}, {Chen}, {Battisti}, {Acharyya}, {Ridolfo}, {Poehler}, {Mably}, {Verma}, {Hayward}, {Kharbanda}, {Poetrodjojo}, {Seibert}, {Rich}, {Madore}, \& {Kewley}}]{Grasha_2022}
{Grasha}, K., {Chen}, Q.~H., {Battisti}, A.~J., {et~al.} 2022, \apj, 929, 118, \dodoi{10.3847/1538-4357/ac5ab2}

\bibitem[{{Grossi} {et~al.}(2020){Grossi}, {Garc{\'\i}a-Benito}, {Cortesi}, {Gon{\c{c}}alves}, {Gon{\c{c}}alves}, {Lopes}, {Men{\'e}ndez-Delmestre}, \& {Telles}}]{Grossi_2020}
{Grossi}, M., {Garc{\'\i}a-Benito}, R., {Cortesi}, A., {et~al.} 2020, \mnras, 498, 1939, \dodoi{10.1093/mnras/staa2382}

\bibitem[{{Hemler} {et~al.}(2021){Hemler}, {Torrey}, {Qi}, {Hernquist}, {Vogelsberger}, {Ma}, {Kewley}, {Nelson}, {Pillepich}, {Pakmor}, \& {Marinacci}}]{Hemler_2021}
{Hemler}, Z.~S., {Torrey}, P., {Qi}, J., {et~al.} 2021, \mnras, 506, 3024, \dodoi{10.1093/mnras/stab1803}

\bibitem[{{Hennebelle} \& {Chabrier}(2011)}]{Hennebelle_2011}
{Hennebelle}, P., \& {Chabrier}, G. 2011, \apjl, 743, L29, \dodoi{10.1088/2041-8205/743/2/L29}

\bibitem[{{Hirai} \& {Saitoh}(2017)}]{Hirai_2017}
{Hirai}, Y., \& {Saitoh}, T.~R. 2017, \apjl, 838, L23, \dodoi{10.3847/2041-8213/aa6799}

\bibitem[{{Hopkins}(2015)}]{Hopkins_2015}
{Hopkins}, P.~F. 2015, \mnras, 450, 53, \dodoi{10.1093/mnras/stv195}

\bibitem[{{Hopkins}(2017)}]{Hopkins_2017}
---. 2017, \mnras, 466, 3387, \dodoi{10.1093/mnras/stw3306}

\bibitem[{{Hopkins} {et~al.}(2014){Hopkins}, {Kere{\v{s}}}, {O{\~n}orbe}, {Faucher-Gigu{\`e}re}, {Quataert}, {Murray}, \& {Bullock}}]{Hopkins_2014}
{Hopkins}, P.~F., {Kere{\v{s}}}, D., {O{\~n}orbe}, J., {et~al.} 2014, \mnras, 445, 581, \dodoi{10.1093/mnras/stu1738}

\bibitem[{{Hopkins} {et~al.}(2018){Hopkins}, {Wetzel}, {Kere{\v{s}}}, {Faucher-Gigu{\`e}re}, {Quataert}, {Boylan-Kolchin}, {Murray}, {Hayward}, {Garrison-Kimmel}, {Hummels}, {Feldmann}, {Torrey}, {Ma}, {Angl{\'e}s-Alc{\'a}zar}, {Su}, {Orr}, {Schmitz}, {Escala}, {Sanderson}, {Grudi{\'c}}, {Hafen}, {Kim}, {Fitts}, {Bullock}, {Wheeler}, {Chan}, {Elbert}, \& {Narayanan}}]{Hopkins_2018}
{Hopkins}, P.~F., {Wetzel}, A., {Kere{\v{s}}}, D., {et~al.} 2018, \mnras, 480, 800, \dodoi{10.1093/mnras/sty1690}

\bibitem[{{Ivey} {et~al.}(2025){Ivey}, {Scholtz}, {Danhaive}, {Koudmani}, {Jones}, {Maiolino}, {Curti}, {D'Eugenio}, {Tacchella}, {Baker}, {Arribas}, {Charlot}, {Eisenstein}, {Ji}, {Laporte}, {Pusk{\'a}s}, {Robertson}, {Sijacki}, \& {Witten}}]{Ivey_2025}
{Ivey}, L.~R., {Scholtz}, J., {Danhaive}, A.~L., {et~al.} 2025, arXiv e-prints, arXiv:2507.14936, \dodoi{10.48550/arXiv.2507.14936}

\bibitem[{{Iwamoto} {et~al.}(1999){Iwamoto}, {Brachwitz}, {Nomoto}, {Kishimoto}, {Umeda}, {Hix}, \& {Thielemann}}]{Iwamoto_1999}
{Iwamoto}, K., {Brachwitz}, F., {Nomoto}, K., {et~al.} 1999, \apjs, 125, 439, \dodoi{10.1086/313278}

\bibitem[{{Izzard} {et~al.}(2004){Izzard}, {Tout}, {Karakas}, \& {Pols}}]{Izzard_2004}
{Izzard}, R.~G., {Tout}, C.~A., {Karakas}, A.~I., \& {Pols}, O.~R. 2004, \mnras, 350, 407, \dodoi{10.1111/j.1365-2966.2004.07446.x}

\bibitem[{{Jijina} \& {Adams}(1996)}]{Jijina_1996}
{Jijina}, J., \& {Adams}, F.~C. 1996, \apj, 462, 874, \dodoi{10.1086/177201}

\bibitem[{{Jones} {et~al.}(2013){Jones}, {Ellis}, {Richard}, \& {Jullo}}]{Jones_2013}
{Jones}, T., {Ellis}, R.~S., {Richard}, J., \& {Jullo}, E. 2013, \apj, 765, 48, \dodoi{10.1088/0004-637X/765/1/48}

\bibitem[{{Jones} {et~al.}(2015){Jones}, {Wang}, {Schmidt}, {Treu}, {Brammer}, {Brada{\v{c}}}, {Dressler}, {Henry}, {Malkan}, {Pentericci}, \& {Trenti}}]{Jones_2015}
{Jones}, T., {Wang}, X., {Schmidt}, K.~B., {et~al.} 2015, \aj, 149, 107, \dodoi{10.1088/0004-6256/149/3/107}

\bibitem[{{Ju} {et~al.}(2025){Ju}, {Wang}, {Jones}, {Bari{\v{s}}i{\'c}}, {Nanayakkara}, {Bundy}, {Faucher-Gigu{\`e}re}, {Feng}, {Glazebrook}, {Henry}, {Malkan}, {Obreschkow}, {Roy}, {Sanders}, {Sun}, {Treu}, \& {Zhou}}]{Ju_2024}
{Ju}, M., {Wang}, X., {Jones}, T., {et~al.} 2025, \apjl, 978, L39, \dodoi{10.3847/2041-8213/ada150}

\bibitem[{{Kannan} {et~al.}(2022){Kannan}, {Garaldi}, {Smith}, {Pakmor}, {Springel}, {Vogelsberger}, \& {Hernquist}}]{Kannan_2022}
{Kannan}, R., {Garaldi}, E., {Smith}, A., {et~al.} 2022, \mnras, 511, 4005, \dodoi{10.1093/mnras/stab3710}

\bibitem[{{Kannan} {et~al.}(2020){Kannan}, {Marinacci}, {Simpson}, {Glover}, \& {Hernquist}}]{Kannan_2020}
{Kannan}, R., {Marinacci}, F., {Simpson}, C.~M., {Glover}, S. C.~O., \& {Hernquist}, L. 2020, \mnras, 491, 2088, \dodoi{10.1093/mnras/stz3078}

\bibitem[{{Kannan} {et~al.}(2019){Kannan}, {Vogelsberger}, {Marinacci}, {McKinnon}, {Pakmor}, \& {Springel}}]{Kannan_2019}
{Kannan}, R., {Vogelsberger}, M., {Marinacci}, F., {et~al.} 2019, \mnras, 485, 117, \dodoi{10.1093/mnras/stz287}

\bibitem[{{Kannan} {et~al.}(2025){Kannan}, {Puchwein}, {Smith}, {Borrow}, {Garaldi}, {Keating}, {Vogelsberger}, {Zier}, {McClymont}, {Shen}, {Popovic}, {Tacchella}, {Hernquist}, \& {Springel}}]{Kannan_2025}
{Kannan}, R., {Puchwein}, E., {Smith}, A., {et~al.} 2025, arXiv e-prints, arXiv:2502.20437, \dodoi{10.48550/arXiv.2502.20437}

\bibitem[{{Karakas}(2010)}]{Karakas_2010}
{Karakas}, A.~I. 2010, \mnras, 403, 1413, \dodoi{10.1111/j.1365-2966.2009.16198.x}

\bibitem[{{Kennicutt}(1998)}]{Kennicutt_1998}
{Kennicutt}, Jr., R.~C. 1998, \araa, 36, 189, \dodoi{10.1146/annurev.astro.36.1.189}

\bibitem[{{Kewley} {et~al.}(2019){Kewley}, {Nicholls}, \& {Sutherland}}]{Kewley_2019}
{Kewley}, L.~J., {Nicholls}, D.~C., \& {Sutherland}, R.~S. 2019, \araa, 57, 511, \dodoi{10.1146/annurev-astro-081817-051832}

\bibitem[{{Knollmann} \& {Knebe}(2009)}]{Knollmann_2009}
{Knollmann}, S.~R., \& {Knebe}, A. 2009, \apjs, 182, 608, \dodoi{10.1088/0067-0049/182/2/608}

\bibitem[{{Kobayashi} {et~al.}(2006){Kobayashi}, {Umeda}, {Nomoto}, {Tominaga}, \& {Ohkubo}}]{Kobayashi_2006}
{Kobayashi}, C., {Umeda}, H., {Nomoto}, K., {Tominaga}, N., \& {Ohkubo}, T. 2006, \apj, 653, 1145, \dodoi{10.1086/508914}

\bibitem[{{Kretschmer} \& {Teyssier}(2020)}]{Kretschmer_2020}
{Kretschmer}, M., \& {Teyssier}, R. 2020, \mnras, 492, 1385, \dodoi{10.1093/mnras/stz3495}

\bibitem[{{Kroupa}(2002)}]{Kroupa_2002}
{Kroupa}, P. 2002, Science, 295, 82, \dodoi{10.1126/science.1067524}

\bibitem[{{Lazar} {et~al.}(2020){Lazar}, {Bullock}, {Boylan-Kolchin}, {Chan}, {Hopkins}, {Graus}, {Wetzel}, {El-Badry}, {Wheeler}, {Straight}, {Kere{\v{s}}}, {Faucher-Gigu{\`e}re}, {Fitts}, \& {Garrison-Kimmel}}]{Lazar_2020}
{Lazar}, A., {Bullock}, J.~S., {Boylan-Kolchin}, M., {et~al.} 2020, \mnras, 497, 2393, \dodoi{10.1093/mnras/staa2101}

\bibitem[{{Leethochawalit} {et~al.}(2016){Leethochawalit}, {Jones}, {Ellis}, {Stark}, {Richard}, {Zitrin}, \& {Auger}}]{Leethochawalit_2016}
{Leethochawalit}, N., {Jones}, T.~A., {Ellis}, R.~S., {et~al.} 2016, \apj, 820, 84, \dodoi{10.3847/0004-637X/820/2/84}

\bibitem[{{Leitherer} {et~al.}(2014){Leitherer}, {Ekstr{\"o}m}, {Meynet}, {Schaerer}, {Agienko}, \& {Levesque}}]{Leitherer_2014}
{Leitherer}, C., {Ekstr{\"o}m}, S., {Meynet}, G., {et~al.} 2014, \apjs, 212, 14, \dodoi{10.1088/0067-0049/212/1/14}

\bibitem[{{Leitherer} {et~al.}(1999){Leitherer}, {Schaerer}, {Goldader}, {Delgado}, {Robert}, {Kune}, {de Mello}, {Devost}, \& {Heckman}}]{Leitherer_1999}
{Leitherer}, C., {Schaerer}, D., {Goldader}, J.~D., {et~al.} 1999, \apjs, 123, 3, \dodoi{10.1086/313233}

\bibitem[{{Li} {et~al.}(2022){Li}, {Wang}, {Cai}, {Shi}, {Fan}, {Zheng}, {Malkan}, {Teplitz}, {Henry}, {Bian}, \& {Colbert}}]{Li_2022}
{Li}, Z., {Wang}, X., {Cai}, Z., {et~al.} 2022, \apjl, 929, L8, \dodoi{10.3847/2041-8213/ac626f}

\bibitem[{{Li} {et~al.}(2025){Li}, {Cai}, {Wang}, {Li}, {Dekel}, {Sarkar}, {Ba{\~n}ados}, {Bian}, {Bhowmick}, {Blecha}, {Bosman}, {Champagne}, {Fan}, {Golden-Marx}, {Jun}, {Li}, {Lin}, {Liu}, {Sun}, {Trebitsch}, {Walter}, {Wang}, {Wu}, {Yang}, {Zhang}, {Zhang}, {Zhuang}, \& {Zou}}]{Li_2025}
{Li}, Z., {Cai}, Z., {Wang}, X., {et~al.} 2025, arXiv e-prints, arXiv:2506.12129, \dodoi{10.48550/arXiv.2506.12129}

\bibitem[{{Looser} {et~al.}(2024){Looser}, {D'Eugenio}, {Maiolino}, {Witstok}, {Sandles}, {Curtis-Lake}, {Chevallard}, {Tacchella}, {Johnson}, {Baker}, {Suess}, {Carniani}, {Ferruit}, {Arribas}, {Bonaventura}, {Bunker}, {Cameron}, {Charlot}, {Curti}, {de Graaff}, {Maseda}, {Rawle}, {Rix}, {Del Pino}, {Smit}, {{\"U}bler}, {Willott}, {Alberts}, {Egami}, {Eisenstein}, {Endsley}, {Hausen}, {Rieke}, {Robertson}, {Shivaei}, {Williams}, {Boyett}, {Chen}, {Ji}, {Jones}, {Kumari}, {Nelson}, {Perna}, {Saxena}, \& {Scholtz}}]{Looser_2024}
{Looser}, T.~J., {D'Eugenio}, F., {Maiolino}, R., {et~al.} 2024, \nat, 629, 53, \dodoi{10.1038/s41586-024-07227-0}

\bibitem[{{Lopez} {et~al.}(2011){Lopez}, {Krumholz}, {Bolatto}, {Prochaska}, \& {Ramirez-Ruiz}}]{Lopez_2011}
{Lopez}, L.~A., {Krumholz}, M.~R., {Bolatto}, A.~D., {Prochaska}, J.~X., \& {Ramirez-Ruiz}, E. 2011, \apj, 731, 91, \dodoi{10.1088/0004-637X/731/2/91}

\bibitem[{Ma {et~al.}(2017)Ma, Hopkins, Feldmann, Torrey, Faucher-Giguère, \& Kereš}]{Ma_2017}
Ma, X., Hopkins, P.~F., Feldmann, R., {et~al.} 2017, MNRAS, 466, 4780, \dodoi{10.1093/mnras/stx034}

\bibitem[{{Ma} {et~al.}(2020){Ma}, {Quataert}, {Wetzel}, {Hopkins}, {Faucher-Gigu{\`e}re}, \& {Kere{\v{s}}}}]{Ma_2020}
{Ma}, X., {Quataert}, E., {Wetzel}, A., {et~al.} 2020, \mnras, 498, 2001, \dodoi{10.1093/mnras/staa2404}

\bibitem[{{Ma} {et~al.}(2018){Ma}, {Hopkins}, {Garrison-Kimmel}, {Faucher-Gigu{\`e}re}, {Quataert}, {Boylan-Kolchin}, {Hayward}, {Feldmann}, \& {Kere{\v{s}}}}]{Ma_2018}
{Ma}, X., {Hopkins}, P.~F., {Garrison-Kimmel}, S., {et~al.} 2018, \mnras, 478, 1694, \dodoi{10.1093/mnras/sty1024}

\bibitem[{{Ma} {et~al.}(2019){Ma}, {Hayward}, {Casey}, {Hopkins}, {Quataert}, {Liang}, {Faucher-Gigu{\`e}re}, {Feldmann}, \& {Kere{\v{s}}}}]{Ma_2019}
{Ma}, X., {Hayward}, C.~C., {Casey}, C.~M., {et~al.} 2019, \mnras, 487, 1844, \dodoi{10.1093/mnras/stz1324}

\bibitem[{{Maiolino} \& {Mannucci}(2019)}]{Maiolino_Mannucci_2019}
{Maiolino}, R., \& {Mannucci}, F. 2019, \aapr, 27, 3, \dodoi{10.1007/s00159-018-0112-2}

\bibitem[{{Mannucci} {et~al.}(2006){Mannucci}, {Della Valle}, \& {Panagia}}]{Mannucci_2006}
{Mannucci}, F., {Della Valle}, M., \& {Panagia}, N. 2006, \mnras, 370, 773, \dodoi{10.1111/j.1365-2966.2006.10501.x}

\bibitem[{{Marigo}(2001)}]{Marigo_2001}
{Marigo}, P. 2001, \aap, 370, 194, \dodoi{10.1051/0004-6361:20000247}

\bibitem[{{Marinacci} {et~al.}(2019){Marinacci}, {Sales}, {Vogelsberger}, {Torrey}, \& {Springel}}]{Marinacci_2019}
{Marinacci}, F., {Sales}, L.~V., {Vogelsberger}, M., {Torrey}, P., \& {Springel}, V. 2019, \mnras, 489, 4233, \dodoi{10.1093/mnras/stz2391}

\bibitem[{{Marszewski} {et~al.}(2025){Marszewski}, {Faucher-Gigu{\`e}re}, {Feldmann}, \& {Sun}}]{Marszewski_2025}
{Marszewski}, A., {Faucher-Gigu{\`e}re}, C.-A., {Feldmann}, R., \& {Sun}, G. 2025, \apjl, 991, L4, \dodoi{10.3847/2041-8213/adf74b}

\bibitem[{{Mason} {et~al.}(2023){Mason}, {Trenti}, \& {Treu}}]{Mason_2023}
{Mason}, C.~A., {Trenti}, M., \& {Treu}, T. 2023, \mnras, 521, 497, \dodoi{10.1093/mnras/stad035}

\bibitem[{{McClymont} {et~al.}(2025{\natexlab{a}}){McClymont}, {Tacchella}, {Smith}, {Kannan}, {Puchwein}, {Borrow}, {Garaldi}, {Keating}, {Vogelsberger}, {Zier}, {Shen}, {Popovic}, \& {Simmonds}}]{McClymont_2025c}
{McClymont}, W., {Tacchella}, S., {Smith}, A., {et~al.} 2025{\natexlab{a}}, arXiv e-prints, arXiv:2503.00106, \dodoi{10.48550/arXiv.2503.00106}

\bibitem[{{McClymont} {et~al.}(2025{\natexlab{b}}){McClymont}, {Tacchella}, {Smith}, {Kannan}, {Puchwein}, {Borrow}, {Garaldi}, {Keating}, {Vogelsberger}, {Zier}, {Shen}, \& {Popovic}}]{McClymont_2025b}
---. 2025{\natexlab{b}}, arXiv e-prints, arXiv:2503.04894, \dodoi{10.48550/arXiv.2503.04894}

\bibitem[{{McClymont} {et~al.}(2025{\natexlab{c}}){McClymont}, {Tacchella}, {Smith}, {Kannan}, {Garaldi}, {Puchwein}, {Isobe}, {Ji}, {Shen}, {Wang}, {Belokurov}, {Borrow}, {D'Eugenio}, {Keating}, {Maiolino}, {Monty}, {Vogelsberger}, \& {Zier}}]{McClymont_2025}
---. 2025{\natexlab{c}}, arXiv e-prints, arXiv:2507.08787, \dodoi{10.48550/arXiv.2507.08787}

\bibitem[{{McKinnon} {et~al.}(2016){McKinnon}, {Torrey}, \& {Vogelsberger}}]{McKinnon_2016}
{McKinnon}, R., {Torrey}, P., \& {Vogelsberger}, M. 2016, \mnras, 457, 3775, \dodoi{10.1093/mnras/stw253}

\bibitem[{{McKinnon} {et~al.}(2017){McKinnon}, {Torrey}, {Vogelsberger}, {Hayward}, \& {Marinacci}}]{McKinnon_2017}
{McKinnon}, R., {Torrey}, P., {Vogelsberger}, M., {Hayward}, C.~C., \& {Marinacci}, F. 2017, \mnras, 468, 1505, \dodoi{10.1093/mnras/stx467}

\bibitem[{{Mostow} {et~al.}(2024){Mostow}, {Torrey}, {Rose}, {Garcia}, {Ahvazi}, {Lisanti}, \& {Kallivayalil}}]{Mostow_2024}
{Mostow}, O., {Torrey}, P., {Rose}, J., {et~al.} 2024, arXiv e-prints, arXiv:2412.09566.
\newblock \doarXiv{2412.09566}

\bibitem[{{Muratov} {et~al.}(2015){Muratov}, {Kere{\v{s}}}, {Faucher-Gigu{\`e}re}, {Hopkins}, {Quataert}, \& {Murray}}]{Muratov_2015}
{Muratov}, A.~L., {Kere{\v{s}}}, D., {Faucher-Gigu{\`e}re}, C.-A., {et~al.} 2015, \mnras, 454, 2691, \dodoi{10.1093/mnras/stv2126}

\bibitem[{{Muratov} {et~al.}(2017){Muratov}, {Kere{\v{s}}}, {Faucher-Gigu{\`e}re}, {Hopkins}, {Ma}, {Angl{\'e}s-Alc{\'a}zar}, {Chan}, {Torrey}, {Hafen}, {Quataert}, \& {Murray}}]{Muratov_2017}
---. 2017, \mnras, 468, 4170, \dodoi{10.1093/mnras/stx667}

\bibitem[{{Narayanan} {et~al.}(2024{\natexlab{a}}){Narayanan}, {Stark}, {Finkelstein}, {Torrey}, {Li}, {Cullen}, {Topping}, {Marinacci}, {Sales}, {Shen}, \& {Vogelsberger}}]{Narayanan_2024}
{Narayanan}, D., {Stark}, D.~P., {Finkelstein}, S.~L., {et~al.} 2024{\natexlab{a}}, arXiv e-prints, arXiv:2408.13312, \dodoi{10.48550/arXiv.2408.13312}

\bibitem[{{Narayanan} {et~al.}(2024{\natexlab{b}}){Narayanan}, {Lower}, {Torrey}, {Brammer}, {Cui}, {Dav{\'e}}, {Iyer}, {Li}, {Lovell}, {Sales}, {Stark}, {Marinacci}, \& {Vogelsberger}}]{Narayanan_2024b}
{Narayanan}, D., {Lower}, S., {Torrey}, P., {et~al.} 2024{\natexlab{b}}, \apj, 961, 73, \dodoi{10.3847/1538-4357/ad0966}

\bibitem[{Nelson {et~al.}(2021)Nelson, Tacchella, Diemer, Leja, Hernquist, Whitaker, Weinberger, Pillepich, Nelson, Terrazas, Nevin, Brammer, Burkhart, Cochrane, van Dokkum, Johnson, Marinacci, Mowla, Pakmor, Skelton, Speagle, Springel, Torrey, Vogelsberger, \& Wuyts}]{Nelson_2021}
Nelson, E.~J., Tacchella, S., Diemer, B., {et~al.} 2021, MNRAS, 508, 219, \dodoi{10.1093/mnras/stab2131}

\bibitem[{{Nomoto} {et~al.}(1997){Nomoto}, {Iwamoto}, {Nakasato}, {Thielemann}, {Brachwitz}, {Tsujimoto}, {Kubo}, \& {Kishimoto}}]{Nomoto_1997}
{Nomoto}, K., {Iwamoto}, K., {Nakasato}, N., {et~al.} 1997, \nphysa, 621, 467, \dodoi{10.1016/S0375-9474(97)00291-1}

\bibitem[{{Nomoto} {et~al.}(2006){Nomoto}, {Tominaga}, {Umeda}, {Kobayashi}, \& {Maeda}}]{Nomoto_2006}
{Nomoto}, K., {Tominaga}, N., {Umeda}, H., {Kobayashi}, C., \& {Maeda}, K. 2006, \nphysa, 777, 424, \dodoi{10.1016/j.nuclphysa.2006.05.008}

\bibitem[{{Orr} {et~al.}(2023){Orr}, {Burkhart}, {Wetzel}, {Hopkins}, {Escala}, {Strom}, {Goldsmith}, {Pineda}, {Hayward}, \& {Loebman}}]{Orr_2023}
{Orr}, M.~E., {Burkhart}, B., {Wetzel}, A., {et~al.} 2023, \mnras, 521, 3708, \dodoi{10.1093/mnras/stad676}

\bibitem[{{Pandya} {et~al.}(2021){Pandya}, {Fielding}, {Angl{\'e}s-Alc{\'a}zar}, {Somerville}, {Bryan}, {Hayward}, {Stern}, {Kim}, {Quataert}, {Forbes}, {Faucher-Gigu{\`e}re}, {Feldmann}, {Hafen}, {Hopkins}, {Kere{\v{s}}}, {Murray}, \& {Wetzel}}]{Pandya_2021}
{Pandya}, V., {Fielding}, D.~B., {Angl{\'e}s-Alc{\'a}zar}, D., {et~al.} 2021, \mnras, 508, 2979, \dodoi{10.1093/mnras/stab2714}

\bibitem[{{P{\'e}roux} \& {Howk}(2020)}]{Peroux_2020}
{P{\'e}roux}, C., \& {Howk}, J.~C. 2020, \araa, 58, 363, \dodoi{10.1146/annurev-astro-021820-120014}

\bibitem[{{Pilkington} {et~al.}(2012){Pilkington}, {Gibson}, {Brook}, {Calura}, {Stinson}, {Thacker}, {Michel-Dansac}, {Bailin}, {Couchman}, {Wadsley}, {Quinn}, \& {Maccio}}]{Pilkington_2012}
{Pilkington}, K., {Gibson}, B.~K., {Brook}, C.~B., {et~al.} 2012, \mnras, 425, 969, \dodoi{10.1111/j.1365-2966.2012.21353.x}

\bibitem[{Pillepich {et~al.}(2018)Pillepich, Springel, Nelson, Genel, Naiman, Pakmor, Hernquist, Torrey, Vogelsberger, Weinberger, \& Marinacci}]{Pillepich_2018a}
Pillepich, A., Springel, V., Nelson, D., {et~al.} 2018, MNRAS, 473, 4077, \dodoi{10.1093/mnras/stx2656}

\bibitem[{{Ploeckinger} {et~al.}(2024){Ploeckinger}, {Nobels}, {Schaller}, \& {Schaye}}]{Ploeckinger_2024}
{Ploeckinger}, S., {Nobels}, F. S.~J., {Schaller}, M., \& {Schaye}, J. 2024, \mnras, 528, 2930, \dodoi{10.1093/mnras/stad3935}

\bibitem[{{Ploeckinger} {et~al.}(2025){Ploeckinger}, {Richings}, {Schaye}, {Trayford}, {Schaller}, \& {Chaikin}}]{Ploeckinger_2025}
{Ploeckinger}, S., {Richings}, A.~J., {Schaye}, J., {et~al.} 2025, \mnras, 543, 891, \dodoi{10.1093/mnras/staf1402}

\bibitem[{{Poetrodjojo} {et~al.}(2019){Poetrodjojo}, {D'Agostino}, {Groves}, {Kewley}, {Ho}, {Rich}, {Madore}, \& {Seibert}}]{Poetrodjojo_2019}
{Poetrodjojo}, H., {D'Agostino}, J.~J., {Groves}, B., {et~al.} 2019, \mnras, 487, 79, \dodoi{10.1093/mnras/stz1241}

\bibitem[{{Pontzen} \& {Governato}(2012)}]{Pontzen_2012}
{Pontzen}, A., \& {Governato}, F. 2012, \mnras, 421, 3464, \dodoi{10.1111/j.1365-2966.2012.20571.x}

\bibitem[{{Portinari} {et~al.}(1998){Portinari}, {Chiosi}, \& {Bressan}}]{Portinari_1998}
{Portinari}, L., {Chiosi}, C., \& {Bressan}, A. 1998, \aap, 334, 505, \dodoi{10.48550/arXiv.astro-ph/9711337}

\bibitem[{{Prantzos} \& {Boissier}(2000)}]{Prantzos_2000}
{Prantzos}, N., \& {Boissier}, S. 2000, \mnras, 313, 338, \dodoi{10.1046/j.1365-8711.2000.03228.x}

\bibitem[{Pérez {et~al.}(2013)Pérez, Cid~Fernandes, González~Delgado, García-Benito, Sánchez, Husemann, Mast, Rodón, Kupko, Backsmann, de~Amorim, van~de Ven, Walcher, Wisotzki, Cortijo-Ferrero, \& {CALIFA Collaboration}}]{Perez_2013}
Pérez, E., Cid~Fernandes, R., González~Delgado, R.~M., {et~al.} 2013, ApJL, 764, L1, \dodoi{10.1088/2041-8205/764/1/L1}

\bibitem[{{Qi} {et~al.}(2025){Qi}, {Garcia}, {Robinson}, {Torrey}, {Moreno}, {Green}, {Evans}, {Hemler}, {Hernquist}, \& {Ellison}}]{Qi_2025}
{Qi}, J., {Garcia}, A.~M., {Robinson}, D., {et~al.} 2025, arXiv e-prints, arXiv:2501.18687, \dodoi{10.48550/arXiv.2501.18687}

\bibitem[{Queyrel {et~al.}(2012)Queyrel, Contini, Kissler-Patig, Epinat, Amram, Garilli, Le~Fèvre, Moultaka, Paioro, Tasca, Tresse, Vergani, López-Sanjuan, \& Perez-Montero}]{Queyrel_2012}
Queyrel, J., Contini, T., Kissler-Patig, M., {et~al.} 2012, A\&A, 539, A93, \dodoi{10.1051/0004-6361/201117718}

\bibitem[{{Quillen} {et~al.}(2005){Quillen}, {Thorndike}, {Cunningham}, {Frank}, {Gutermuth}, {Blackman}, {Pipher}, \& {Ridge}}]{Quillen_2005}
{Quillen}, A.~C., {Thorndike}, S.~L., {Cunningham}, A., {et~al.} 2005, \apj, 632, 941, \dodoi{10.1086/444410}

\bibitem[{{Read} \& {Gilmore}(2005)}]{Read_2005}
{Read}, J.~I., \& {Gilmore}, G. 2005, \mnras, 356, 107, \dodoi{10.1111/j.1365-2966.2004.08424.x}

\bibitem[{{Rodr{\'\i}guez Del Pino} {et~al.}(2024){Rodr{\'\i}guez Del Pino}, {Perna}, {Arribas}, {D'Eugenio}, {Lamperti}, {P{\'e}rez-Gonz{\'a}lez}, {{\"U}bler}, {Bunker}, {Carniani}, {Charlot}, {Maiolino}, {Willott}, {B{\"o}ker}, {Chevallard}, {Cresci}, {Curti}, {Jones}, {Parlanti}, {Scholtz}, \& {Venturi}}]{Rodriguez_Del_Pino_2024}
{Rodr{\'\i}guez Del Pino}, B., {Perna}, M., {Arribas}, S., {et~al.} 2024, \aap, 684, A187, \dodoi{10.1051/0004-6361/202348057}

\bibitem[{{Rosdahl} {et~al.}(2013){Rosdahl}, {Blaizot}, {Aubert}, {Stranex}, \& {Teyssier}}]{Rosdahl_2013}
{Rosdahl}, J., {Blaizot}, J., {Aubert}, D., {Stranex}, T., \& {Teyssier}, R. 2013, \mnras, 436, 2188, \dodoi{10.1093/mnras/stt1722}

\bibitem[{{Rosdahl} \& {Teyssier}(2015)}]{Rosdahl_2015}
{Rosdahl}, J., \& {Teyssier}, R. 2015, \mnras, 449, 4380, \dodoi{10.1093/mnras/stv567}

\bibitem[{{Rupke} {et~al.}(2010{\natexlab{a}}){Rupke}, {Kewley}, \& {Barnes}}]{Rupke_2010a}
{Rupke}, D. S.~N., {Kewley}, L.~J., \& {Barnes}, J.~E. 2010{\natexlab{a}}, \apjl, 710, L156, \dodoi{10.1088/2041-8205/710/2/L156}

\bibitem[{{Rupke} {et~al.}(2010{\natexlab{b}}){Rupke}, {Kewley}, \& {Chien}}]{Rupke_2010b}
{Rupke}, D. S.~N., {Kewley}, L.~J., \& {Chien}, L.~H. 2010{\natexlab{b}}, \apj, 723, 1255, \dodoi{10.1088/0004-637X/723/2/1255}

\bibitem[{{Samuel} {et~al.}(2020){Samuel}, {Wetzel}, {Tollerud}, {Garrison-Kimmel}, {Loebman}, {El-Badry}, {Hopkins}, {Boylan-Kolchin}, {Faucher-Gigu{\`e}re}, {Bullock}, {Benincasa}, \& {Bailin}}]{Samuel_2020}
{Samuel}, J., {Wetzel}, A., {Tollerud}, E., {et~al.} 2020, \mnras, 491, 1471, \dodoi{10.1093/mnras/stz3054}

\bibitem[{{S{\'a}nchez} {et~al.}(2012){S{\'a}nchez}, {Rosales-Ortega}, {Marino}, {Iglesias-P{\'a}ramo}, {V{\'\i}lchez}, {Kennicutt}, {D{\'\i}az}, {Mast}, {Monreal-Ibero}, {Garc{\'\i}a-Benito}, {Bland-Hawthorn}, {P{\'e}rez}, {Gonz{\'a}lez Delgado}, {Husemann}, {L{\'o}pez-S{\'a}nchez}, {Cid Fernandes}, {Kehrig}, {Walcher}, {Gil de Paz}, \& {Ellis}}]{Sanchez_2012}
{S{\'a}nchez}, S.~F., {Rosales-Ortega}, F.~F., {Marino}, R.~A., {et~al.} 2012, \aap, 546, A2, \dodoi{10.1051/0004-6361/201219578}

\bibitem[{{S{\'a}nchez} {et~al.}(2013){S{\'a}nchez}, {Rosales-Ortega}, {Jungwiert}, {Iglesias-P{\'a}ramo}, {V{\'{\i}}lchez}, {Marino}, {Walcher}, {Husemann}, {Mast}, {Monreal-Ibero}, {Cid Fernandes}, {P{\'e}rez}, {Gonz{\'a}lez Delgado}, {Garc{\'{\i}}a-Benito}, {Galbany}, {van de Ven}, {Jahnke}, {Flores}, {Bland-Hawthorn}, {L{\'o}pez-S{\'a}nchez}, {Stanishev}, {Miralles-Caballero}, {D{\'{\i}}az}, {S{\'a}nchez-Blazquez}, {Moll{\'a}}, {Gallazzi}, {Papaderos}, {Gomes}, {Gruel}, {P{\'e}rez}, {Ruiz-Lara}, {Florido}, {de Lorenzo-C{\'a}ceres}, {Mendez-Abreu}, {Kehrig}, {Roth}, {Ziegler}, {Alves}, {Wisotzki}, {Kupko}, {Quirrenbach}, {Bomans}, \& {Califa Collaboration}}]{Sanchez_2013}
{S{\'a}nchez}, S.~F., {Rosales-Ortega}, F.~F., {Jungwiert}, B., {et~al.} 2013, \aap, 554, A58, \dodoi{10.1051/0004-6361/201220669}

\bibitem[{{S{\'a}nchez} {et~al.}(2014){S{\'a}nchez}, {Rosales-Ortega}, {Iglesias-P{\'a}ramo}, {Moll{\'a}}, {Barrera-Ballesteros}, {Marino}, {P{\'e}rez}, {S{\'a}nchez-Blazquez}, {Gonz{\'a}lez Delgado}, {Cid Fernandes}, {de Lorenzo-C{\'a}ceres}, {Mendez-Abreu}, {Galbany}, {Falcon-Barroso}, {Miralles-Caballero}, {Husemann}, {Garc{\'{\i}}a-Benito}, {Mast}, {Walcher}, {Gil de Paz}, {Garc{\'{\i}}a-Lorenzo}, {Jungwiert}, {V{\'{\i}}lchez}, {J{\'{\i}}lkov{\'a}}, {Lyubenova}, {Cortijo-Ferrero}, {D{\'{\i}}az}, {Wisotzki}, {M{\'a}rquez}, {Bland-Hawthorn}, {Ellis}, {van de Ven}, {Jahnke}, {Papaderos}, {Gomes}, {Mendoza}, \& {L{\'o}pez-S{\'a}nchez}}]{Sanchez_2014}
{S{\'a}nchez}, S.~F., {Rosales-Ortega}, F.~F., {Iglesias-P{\'a}ramo}, J., {et~al.} 2014, \aap, 563, A49, \dodoi{10.1051/0004-6361/201322343}

\bibitem[{{S{\'a}nchez-Menguiano} {et~al.}(2016){S{\'a}nchez-Menguiano}, {S{\'a}nchez}, {P{\'e}rez}, {Garc{\'\i}a-Benito}, {Husemann}, {Mast}, {Mendoza}, {Ruiz-Lara}, {Ascasibar}, {Bland-Hawthorn}, {Cavichia}, {D{\'\i}az}, {Florido}, {Galbany}, {G{\'o}nzalez Delgado}, {Kehrig}, {Marino}, {M{\'a}rquez}, {Masegosa}, {M{\'e}ndez-Abreu}, {Moll{\'a}}, {Del Olmo}, {P{\'e}rez}, {S{\'a}nchez-Bl{\'a}zquez}, {Stanishev}, {Walcher}, {L{\'o}pez-S{\'a}nchez}, \& {CALIFA Collaboration}}]{Sanchez_Menguiano_2016}
{S{\'a}nchez-Menguiano}, L., {S{\'a}nchez}, S.~F., {P{\'e}rez}, I., {et~al.} 2016, \aap, 587, A70, \dodoi{10.1051/0004-6361/201527450}

\bibitem[{Schaye \& Dalla~Vecchia(2008)}]{Schaye_DallaVechhia_2008}
Schaye, J., \& Dalla~Vecchia, C. 2008, MNRAS, 383, 1210, \dodoi{10.1111/j.1365-2966.2007.12639.x}

\bibitem[{{Schaye} {et~al.}(2015){Schaye}, {Crain}, {Bower}, {Furlong}, {Schaller}, {Theuns}, {Dalla Vecchia}, {Frenk}, {McCarthy}, {Helly}, {Jenkins}, {Rosas-Guevara}, {White}, {Baes}, {Booth}, {Camps}, {Navarro}, {Qu}, {Rahmati}, {Sawala}, {Thomas}, \& {Trayford}}]{Schaye_2015}
{Schaye}, J., {Crain}, R.~A., {Bower}, R.~G., {et~al.} 2015, \mnras, 446, 521, \dodoi{10.1093/mnras/stu2058}

\bibitem[{{Schaye} {et~al.}(2025){Schaye}, {Chaikin}, {Schaller}, {Ploeckinger}, {Hu{\v{s}}ko}, {McGibbon}, {Trayford}, {Ben{\'\i}tez-Llambay}, {Correa}, {Frenk}, {Richings}, {Forouhar Moreno}, {Bah{\'e}}, {Borrow}, {Durrant}, {Gebek}, {Helly}, {Jenkins}, {Lacey}, {Ludlow}, \& {Nobels}}]{Schaye_2025}
{Schaye}, J., {Chaikin}, E., {Schaller}, M., {et~al.} 2025, arXiv e-prints, arXiv:2508.21126, \dodoi{10.48550/arXiv.2508.21126}

\bibitem[{{Schmidt}(1959)}]{Schmidt_1959}
{Schmidt}, M. 1959, \apj, 129, 243, \dodoi{10.1086/146614}

\bibitem[{{Schmidt} {et~al.}(2006){Schmidt}, {Niemeyer}, {Hillebrandt}, \& {R{\"o}pke}}]{Schmidt_2006}
{Schmidt}, W., {Niemeyer}, J.~C., {Hillebrandt}, W., \& {R{\"o}pke}, F.~K. 2006, \aap, 450, 283, \dodoi{10.1051/0004-6361:20053618}

\bibitem[{Searle(1971)}]{Searle_1971}
Searle, L. 1971, ApJ, 168, 327, \dodoi{10.1086/151090}

\bibitem[{{Shen} {et~al.}(2023){Shen}, {Vogelsberger}, {Boylan-Kolchin}, {Tacchella}, \& {Kannan}}]{Shen_2023}
{Shen}, X., {Vogelsberger}, M., {Boylan-Kolchin}, M., {Tacchella}, S., \& {Kannan}, R. 2023, \mnras, 525, 3254, \dodoi{10.1093/mnras/stad2508}

\bibitem[{{Shen} {et~al.}(2024{\natexlab{a}}){Shen}, {Vogelsberger}, {Boylan-Kolchin}, {Tacchella}, \& {Naidu}}]{Shen_2024}
{Shen}, X., {Vogelsberger}, M., {Boylan-Kolchin}, M., {Tacchella}, S., \& {Naidu}, R.~P. 2024{\natexlab{a}}, \mnras, 533, 3923, \dodoi{10.1093/mnras/stae1932}

\bibitem[{{Shen} {et~al.}(2025{\natexlab{a}}){Shen}, {Zier}, {Vogelsberger}, {Boylan-Kolchin}, {Hernquist}, {Tacchella}, \& {Naidu}}]{Shen_2025b}
{Shen}, X., {Zier}, O., {Vogelsberger}, M., {et~al.} 2025{\natexlab{a}}, arXiv e-prints, arXiv:2509.19427, \dodoi{10.48550/arXiv.2509.19427}

\bibitem[{{Shen} {et~al.}(2024{\natexlab{b}}){Shen}, {Vogelsberger}, {Borrow}, {Hu}, {Erickson}, {Kannan}, {Smith}, {Garaldi}, {Hernquist}, {Morishita}, {Tacchella}, {Zier}, {Sun}, {Eilers}, \& {Wang}}]{Shen_2024b}
{Shen}, X., {Vogelsberger}, M., {Borrow}, J., {et~al.} 2024{\natexlab{b}}, \mnras, 534, 1433, \dodoi{10.1093/mnras/stae2156}

\bibitem[{{Shen} {et~al.}(2025{\natexlab{b}}){Shen}, {Kannan}, {Puchwein}, {Smith}, {Vogelsberger}, {Borrow}, {Garaldi}, {Keating}, {Zier}, {McClymont}, {Tacchella}, {Wang}, \& {Hernquist}}]{Shen_2025}
{Shen}, X., {Kannan}, R., {Puchwein}, E., {et~al.} 2025{\natexlab{b}}, arXiv e-prints, arXiv:2503.01949, \dodoi{10.48550/arXiv.2503.01949}

\bibitem[{{Simons} {et~al.}(2021){Simons}, {Papovich}, {Momcheva}, {Trump}, {Brammer}, {Estrada-Carpenter}, {Backhaus}, {Cleri}, {Finkelstein}, {Giavalisco}, {Ji}, {Jung}, {Matharu}, \& {Weiner}}]{Simons_2021}
{Simons}, R.~C., {Papovich}, C., {Momcheva}, I., {et~al.} 2021, \apj, 923, 203, \dodoi{10.3847/1538-4357/ac28f4}

\bibitem[{{Smith} {et~al.}(2022){Smith}, {Kannan}, {Garaldi}, {Vogelsberger}, {Pakmor}, {Springel}, \& {Hernquist}}]{Smith_2022}
{Smith}, A., {Kannan}, R., {Garaldi}, E., {et~al.} 2022, \mnras, 512, 3243, \dodoi{10.1093/mnras/stac713}

\bibitem[{{Sparre} {et~al.}(2017){Sparre}, {Hayward}, {Feldmann}, {Faucher-Gigu{\`e}re}, {Muratov}, {Kere{\v{s}}}, \& {Hopkins}}]{Sparre_2017}
{Sparre}, M., {Hayward}, C.~C., {Feldmann}, R., {et~al.} 2017, \mnras, 466, 88, \dodoi{10.1093/mnras/stw3011}

\bibitem[{{Sparre} {et~al.}(2015){Sparre}, {Hayward}, {Springel}, {Vogelsberger}, {Genel}, {Torrey}, {Nelson}, {Sijacki}, \& {Hernquist}}]{Sparre_2015}
{Sparre}, M., {Hayward}, C.~C., {Springel}, V., {et~al.} 2015, \mnras, 447, 3548, \dodoi{10.1093/mnras/stu2713}

\bibitem[{{Springel}(2010)}]{Springel_2010}
{Springel}, V. 2010, \mnras, 401, 791, \dodoi{10.1111/j.1365-2966.2009.15715.x}

\bibitem[{{Springel} \& {Hernquist}(2003)}]{Springel_Hernquist_2003}
{Springel}, V., \& {Hernquist}, L. 2003, \mnras, 339, 289, \dodoi{10.1046/j.1365-8711.2003.06206.x}

\bibitem[{{Springel} {et~al.}(2021){Springel}, {Pakmor}, {Zier}, \& {Reinecke}}]{Springel_2021}
{Springel}, V., {Pakmor}, R., {Zier}, O., \& {Reinecke}, M. 2021, \mnras, 506, 2871, \dodoi{10.1093/mnras/stab1855}

\bibitem[{Springel {et~al.}(2001)Springel, White, \& Hernquist}]{Springel_2001}
Springel, V., White, M., \& Hernquist, L. 2001, ApJ, 549, 681, \dodoi{10.1086/319473}

\bibitem[{{Stanghellini} {et~al.}(2015){Stanghellini}, {Magrini}, \& {Casasola}}]{Stanghellini_2015}
{Stanghellini}, L., {Magrini}, L., \& {Casasola}, V. 2015, \apj, 812, 39, \dodoi{10.1088/0004-637X/812/1/39}

\bibitem[{{Stinson} {et~al.}(2010){Stinson}, {Bailin}, {Couchman}, {Wadsley}, {Shen}, {Nickerson}, {Brook}, \& {Quinn}}]{Stinson_2010}
{Stinson}, G.~S., {Bailin}, J., {Couchman}, H., {et~al.} 2010, \mnras, 408, 812, \dodoi{10.1111/j.1365-2966.2010.17187.x}

\bibitem[{{Su} {et~al.}(2017){Su}, {Hopkins}, {Hayward}, {Faucher-Gigu{\`e}re}, {Kere{\v{s}}}, {Ma}, \& {Robles}}]{Su_2017}
{Su}, K.-Y., {Hopkins}, P.~F., {Hayward}, C.~C., {et~al.} 2017, \mnras, 471, 144, \dodoi{10.1093/mnras/stx1463}

\bibitem[{{Sun} {et~al.}(2023){Sun}, {Faucher-Gigu{\`e}re}, {Hayward}, {Shen}, {Wetzel}, \& {Cochrane}}]{Sun_2023}
{Sun}, G., {Faucher-Gigu{\`e}re}, C.-A., {Hayward}, C.~C., {et~al.} 2023, \apjl, 955, L35, \dodoi{10.3847/2041-8213/acf85a}

\bibitem[{{Sun} {et~al.}(2024{\natexlab{a}}){Sun}, {Ho}, {Zhuang}, {Ma}, {Chen}, \& {Li}}]{SunW_2024}
{Sun}, W., {Ho}, L.~C., {Zhuang}, M.-Y., {et~al.} 2024{\natexlab{a}}, \apj, 960, 104, \dodoi{10.3847/1538-4357/acf1f6}

\bibitem[{{Sun} {et~al.}(2024{\natexlab{b}}){Sun}, {Wang}, {Ma}, {Wang}, {Wetzel}, {Faucher-Gigu{\`e}re}, {Hopkins}, {Kere{\v{s}}}, {Graf}, {Marszewski}, {Stern}, {Sun}, {Sun}, \& {Thyme}}]{Sun_2024}
{Sun}, X., {Wang}, X., {Ma}, X., {et~al.} 2024{\natexlab{b}}, arXiv e-prints, arXiv:2409.09290, \dodoi{10.48550/arXiv.2409.09290}

\bibitem[{Swinbank {et~al.}(2012)Swinbank, Sobral, Smail, Geach, Best, McCarthy, Crain, \& Theuns}]{Swinbank_2012}
Swinbank, A.~M., Sobral, D., Smail, I., {et~al.} 2012, MNRAS, 426, 935, \dodoi{10.1111/j.1365-2966.2012.21774.x}

\bibitem[{{Tapia-Contreras} {et~al.}(2025){Tapia-Contreras}, {Tissera}, {Sillero}, {Gonzalez-Jara}, {Casanueva-Villarreal}, {Pedrosa}, {Bignone}, {Padilla}, \& {Dom{\'\i}nguez-Tenreiro}}]{Tapia_2025}
{Tapia-Contreras}, B., {Tissera}, P.~B., {Sillero}, E., {et~al.} 2025, \aap, 700, A69, \dodoi{10.1051/0004-6361/202554013}

\bibitem[{{Teyssier}(2002)}]{Teyssier_2002}
{Teyssier}, R. 2002, \aap, 385, 337, \dodoi{10.1051/0004-6361:20011817}

\bibitem[{{Tissera} {et~al.}(2019){Tissera}, {Rosas-Guevara}, {Bower}, {Crain}, {del P Lagos}, {Schaller}, {Schaye}, \& {Theuns}}]{Tissera_2019}
{Tissera}, P.~B., {Rosas-Guevara}, Y., {Bower}, R.~G., {et~al.} 2019, \mnras, 482, 2208, \dodoi{10.1093/mnras/sty2817}

\bibitem[{{Tissera} {et~al.}(2022){Tissera}, {Rosas-Guevara}, {Sillero}, {Pedrosa}, {Theuns}, \& {Bignone}}]{Tissera_2022}
{Tissera}, P.~B., {Rosas-Guevara}, Y., {Sillero}, E., {et~al.} 2022, \mnras, 511, 1667, \dodoi{10.1093/mnras/stab3644}

\bibitem[{{Torrey} {et~al.}(2012){Torrey}, {Cox}, {Kewley}, \& {Hernquist}}]{Torrey_2012}
{Torrey}, P., {Cox}, T.~J., {Kewley}, L., \& {Hernquist}, L. 2012, \apj, 746, 108, \dodoi{10.1088/0004-637X/746/1/108}

\bibitem[{{Torrey} {et~al.}(2017){Torrey}, {Hopkins}, {Faucher-Gigu{\`e}re}, {Vogelsberger}, {Quataert}, {Kere{\v{s}}}, \& {Murray}}]{Torrey_2017}
{Torrey}, P., {Hopkins}, P.~F., {Faucher-Gigu{\`e}re}, C.-A., {et~al.} 2017, \mnras, 467, 2301, \dodoi{10.1093/mnras/stx254}

\bibitem[{{Torrey} {et~al.}(2019){Torrey}, {Vogelsberger}, {Marinacci}, {Pakmor}, {Springel}, {Nelson}, {Naiman}, {Pillepich}, {Genel}, {Weinberger}, \& {Hernquist}}]{Torrey_2019}
{Torrey}, P., {Vogelsberger}, M., {Marinacci}, F., {et~al.} 2019, \mnras, 484, 5587, \dodoi{10.1093/mnras/stz243}

\bibitem[{{Tripodi} {et~al.}(2024){Tripodi}, {D'Eugenio}, {Maiolino}, {Curti}, {Scholtz}, {Tacchella}, {Marconcini}, {Bunker}, {Trussler}, {Cameron}, {Arribas}, {Baker}, {Brada{\v{c}}}, {Carniani}, {Charlot}, {Ji}, {Ji}, {Robertson}, {{\"U}bler}, {Venturi}, {Willmer}, \& {Witstok}}]{Tripodi_2024}
{Tripodi}, R., {D'Eugenio}, F., {Maiolino}, R., {et~al.} 2024, \aap, 692, A184, \dodoi{10.1051/0004-6361/202449980}

\bibitem[{{Troncoso} {et~al.}(2014){Troncoso}, {Maiolino}, {Sommariva}, {Cresci}, {Mannucci}, {Marconi}, {Meneghetti}, {Grazian}, {Cimatti}, {Fontana}, {Nagao}, \& {Pentericci}}]{Troncoso_2014}
{Troncoso}, P., {Maiolino}, R., {Sommariva}, V., {et~al.} 2014, \aap, 563, A58, \dodoi{10.1051/0004-6361/201322099}

\bibitem[{{Tumlinson} {et~al.}(2017){Tumlinson}, {Peeples}, \& {Werk}}]{Tumlinson_2017}
{Tumlinson}, J., {Peeples}, M.~S., \& {Werk}, J.~K. 2017, \araa, 55, 389, \dodoi{10.1146/annurev-astro-091916-055240}

\bibitem[{{Val{\'e}} {et~al.}(2025){Val{\'e}}, {Lara-L{\'o}pez}, {Valerdi}, {Zinchenko}, {O'Sullivan}, {Pilyugin}, {Cepa}, {Casasola}, {De Rossi}, {Dib}, {Fritz}, {Gallego}, {Gardu{\~n}o}, {L{\'o}pez-Cruz}, {Tailor}, \& {Zaragoza-Cardiel}}]{Vale_2025}
{Val{\'e}}, G., {Lara-L{\'o}pez}, M.~A., {Valerdi}, M., {et~al.} 2025, \aap, 701, A226, \dodoi{10.1051/0004-6361/202553818}

\bibitem[{{Vallini} {et~al.}(2024){Vallini}, {Witstok}, {Sommovigo}, {Pallottini}, {Ferrara}, {Carniani}, {Kohandel}, {Smit}, {Gallerani}, \& {Gruppioni}}]{Vallini_2024}
{Vallini}, L., {Witstok}, J., {Sommovigo}, L., {et~al.} 2024, \mnras, 527, 10, \dodoi{10.1093/mnras/stad3150}

\bibitem[{{van den Hoek} \& {Groenewegen}(1997)}]{van_den_hoek_1997}
{van den Hoek}, L.~B., \& {Groenewegen}, M.~A.~T. 1997, \aaps, 123, 305, \dodoi{10.1051/aas:1997162}

\bibitem[{{Venturi} {et~al.}(2024){Venturi}, {Carniani}, {Parlanti}, {Kohandel}, {Curti}, {Pallottini}, {Vallini}, {Arribas}, {Bunker}, {Cameron}, {Castellano}, {Ferrara}, {Fontana}, {Gallerani}, {Gelli}, {Maiolino}, {Ntormousi}, {Pacifici}, {Pentericci}, {Salvadori}, \& {Vanzella}}]{Venturi_2024}
{Venturi}, G., {Carniani}, S., {Parlanti}, E., {et~al.} 2024, arXiv e-prints, arXiv:2403.03977, \dodoi{10.48550/arXiv.2403.03977}

\bibitem[{{Vogelsberger} {et~al.}(2013){Vogelsberger}, {Genel}, {Sijacki}, {Torrey}, {Springel}, \& {Hernquist}}]{Vogelsberger_2013}
{Vogelsberger}, M., {Genel}, S., {Sijacki}, D., {et~al.} 2013, \mnras, 436, 3031, \dodoi{10.1093/mnras/stt1789}

\bibitem[{{Vogelsberger} {et~al.}(2020){Vogelsberger}, {Marinacci}, {Torrey}, \& {Puchwein}}]{Vogelsberger_2020}
{Vogelsberger}, M., {Marinacci}, F., {Torrey}, P., \& {Puchwein}, E. 2020, Nature Reviews Physics, 2, 42, \dodoi{10.1038/s42254-019-0127-2}

\bibitem[{{Vogelsberger} {et~al.}(2014){Vogelsberger}, {Genel}, {Springel}, {Torrey}, {Sijacki}, {Xu}, {Snyder}, {Nelson}, \& {Hernquist}}]{Vogelsberger_2014a}
{Vogelsberger}, M., {Genel}, S., {Springel}, V., {et~al.} 2014, \mnras, 444, 1518, \dodoi{10.1093/mnras/stu1536}

\bibitem[{{Wang} {et~al.}(2017){Wang}, {Jones}, {Treu}, {Morishita}, {Abramson}, {Brammer}, {Huang}, {Malkan}, {Schmidt}, {Fontana}, {Grillo}, {Henry}, {Karman}, {Kelly}, {Mason}, {Mercurio}, {Rosati}, {Sharon}, {Trenti}, \& {Vulcani}}]{Wang_2017}
{Wang}, X., {Jones}, T.~A., {Treu}, T., {et~al.} 2017, \apj, 837, 89, \dodoi{10.3847/1538-4357/aa603c}

\bibitem[{{Wang} {et~al.}(2019){Wang}, {Jones}, {Treu}, {Hirtenstein}, {Brammer}, {Daddi}, {Meng}, {Morishita}, {Abramson}, {Henry}, {Peng}, {Schmidt}, {Sharon}, {Trenti}, \& {Vulcani}}]{Wang_2019}
---. 2019, \apj, 882, 94, \dodoi{10.3847/1538-4357/ab3861}

\bibitem[{{Wang} {et~al.}(2020){Wang}, {Jones}, {Treu}, {Daddi}, {Brammer}, {Sharon}, {Morishita}, {Abramson}, {Colbert}, {Henry}, {Hopkins}, {Malkan}, {Schmidt}, {Teplitz}, \& {Vulcani}}]{Wang_2020}
---. 2020, \apj, 900, 183, \dodoi{10.3847/1538-4357/abacce}

\bibitem[{{Wang} {et~al.}(2022){Wang}, {Jones}, {Vulcani}, {Treu}, {Morishita}, {Roberts-Borsani}, {Malkan}, {Henry}, {Brammer}, {Strait}, {Brada{\v{c}}}, {Boyett}, {Calabr{\`o}}, {Castellano}, {Fontana}, {Glazebrook}, {Kelly}, {Leethochawalit}, {Marchesini}, {Santini}, {Trenti}, \& {Yang}}]{Wang_2022}
{Wang}, X., {Jones}, T., {Vulcani}, B., {et~al.} 2022, \apjl, 938, L16, \dodoi{10.3847/2041-8213/ac959e}

\bibitem[{{Wang} {et~al.}(2025){Wang}, {Shen}, {Vogelsberger}, {Li}, {Kannan}, {Puchwein}, {Smith}, {Borrow}, {Garaldi}, {Keating}, {Zier}, {McClymont}, {Tacchella}, {Ni}, \& {Hernquist}}]{Wang_2025}
{Wang}, Z., {Shen}, X., {Vogelsberger}, M., {et~al.} 2025, arXiv e-prints, arXiv:2505.05554, \dodoi{10.48550/arXiv.2505.05554}

\bibitem[{Weinberger {et~al.}(2017)Weinberger, Springel, Hernquist, Pillepich, Marinacci, Pakmor, Nelson, Genel, Vogelsberger, Naiman, \& Torrey}]{Weinberger_2017}
Weinberger, R., Springel, V., Hernquist, L., {et~al.} 2017, MNRAS, 465, 3291, \dodoi{10.1093/mnras/stw2944}

\bibitem[{{Weinberger} {et~al.}(2018){Weinberger}, {Springel}, {Pakmor}, {Nelson}, {Genel}, {Pillepich}, {Vogelsberger}, {Marinacci}, {Naiman}, {Torrey}, \& {Hernquist}}]{Weinberger_2018}
{Weinberger}, R., {Springel}, V., {Pakmor}, R., {et~al.} 2018, \mnras, 479, 4056, \dodoi{10.1093/mnras/sty1733}

\bibitem[{{Wetzel} {et~al.}(2023){Wetzel}, {Hayward}, {Sanderson}, {Ma}, {Angl{\'e}s-Alc{\'a}zar}, {Feldmann}, {Chan}, {El-Badry}, {Wheeler}, {Garrison-Kimmel}, {Nikakhtar}, {Panithanpaisal}, {Arora}, {Gurvich}, {Samuel}, {Sameie}, {Pandya}, {Hafen}, {Hummels}, {Loebman}, {Boylan-Kolchin}, {Bullock}, {Faucher-Gigu{\`e}re}, {Kere{\v{s}}}, {Quataert}, \& {Hopkins}}]{Wetzel_2023}
{Wetzel}, A., {Hayward}, C.~C., {Sanderson}, R.~E., {et~al.} 2023, \apjs, 265, 44, \dodoi{10.3847/1538-4365/acb99a}

\bibitem[{{Wetzel} {et~al.}(2025){Wetzel}, {Samuel}, {Gandhi}, {Ponnada}, {Su}, {Arora}, {Angles-Alcazar}, {Hayward}, {Sanderson}, {Feldmann}, {Cochrane}, {Nikakhtar}, {Panithanpaisal}, {Benavides}, {Pandya}, {Grudic}, {Hummels}, {Gurvich}, {Hafen}, {Ma}, {Garrison-Kimmel}, {Sameie}, {Chan}, {El-Badry}, {Necib}, {Loebman}, {Wellons}, {Robles}, {Wheeler}, {Moreno}, {Stern}, {Boylan-Kolchin}, {Bullock}, {Faucher-Giguere}, {Keres}, {Quataert}, \& {Hopkins}}]{Wetzel_2025}
{Wetzel}, A., {Samuel}, J., {Gandhi}, P.~J., {et~al.} 2025, arXiv e-prints, arXiv:2508.06608, \dodoi{10.48550/arXiv.2508.06608}

\bibitem[{{Wetzel} {et~al.}(2016){Wetzel}, {Hopkins}, {Kim}, {Faucher-Gigu{\`e}re}, {Kere{\v{s}}}, \& {Quataert}}]{Wetzel_2016}
{Wetzel}, A.~R., {Hopkins}, P.~F., {Kim}, J.-h., {et~al.} 2016, \apjl, 827, L23, \dodoi{10.3847/2041-8205/827/2/L23}

\bibitem[{{Witten} {et~al.}(2025){Witten}, {McClymont}, {Laporte}, {Roberts-Borsani}, {Sijacki}, {Tacchella}, {Simmonds}, {Katz}, {Ellis}, {Witstok}, {Maiolino}, {Ji}, {Hayes}, {Looser}, \& {D'Eugenio}}]{Witten_2025}
{Witten}, C., {McClymont}, W., {Laporte}, N., {et~al.} 2025, \mnras, 537, 112, \dodoi{10.1093/mnras/staf001}

\bibitem[{{Wuyts} {et~al.}(2016){Wuyts}, {Wisnioski}, {Fossati}, {F{\"o}rster Schreiber}, {Genzel}, {Davies}, {Mendel}, {Naab}, {R{\"o}ttgers}, {Wilman}, {Wuyts}, {Bandara}, {Beifiori}, {Belli}, {Bender}, {Brammer}, {Burkert}, {Chan}, {Galametz}, {Kulkarni}, {Lang}, {Lutz}, {Momcheva}, {Nelson}, {Rosario}, {Saglia}, {Seitz}, {Tacconi}, {Tadaki}, {{\"U}bler}, \& {van Dokkum}}]{Wuyts_2016}
{Wuyts}, E., {Wisnioski}, E., {Fossati}, M., {et~al.} 2016, \apj, 827, 74, \dodoi{10.3847/0004-637X/827/1/74}

\bibitem[{{Yorke} {et~al.}(1989){Yorke}, {Tenorio-Tagle}, {Bodenheimer}, \& {Rozyczka}}]{Yorke_1989}
{Yorke}, H.~W., {Tenorio-Tagle}, G., {Bodenheimer}, P., \& {Rozyczka}, M. 1989, \aap, 216, 207

\bibitem[{{Yuan} {et~al.}(2013){Yuan}, {Kewley}, \& {Rich}}]{Yuan_2013}
{Yuan}, T.~T., {Kewley}, L.~J., \& {Rich}, J. 2013, \apj, 767, 106, \dodoi{10.1088/0004-637X/767/2/106}

\end{thebibliography}
\bibliographystyle{aasjournal}

\end{document}